\documentclass[10pt]{iopart}

\usepackage{graphicx}
\usepackage{iopams}
\usepackage{hyperref}
\usepackage{cite}
\usepackage{etoolbox}
\usepackage{dsfont}

\hypersetup{
    colorlinks,
    linkcolor=blue,
    citecolor=blue,
    urlcolor=blue,
    linktocpage=true
}

\makeatletter
\newcommand{\mainmatter}{%
  \setcounter{footnote}{0}
  \patchcmd{\@makefntext}{\fnsymbol}{\arabic}{}{}
  \patchcmd{\@thefnmark}{\fnsymbol}{\arabic}{}{}
  \def\@makefnmark{\textsuperscript{\arabic{footnote}}}
  \def\@makefntext{\textsuperscript{\arabic{footnote}}}
}
\makeatother

\newcommand{\nio}{Na$_2$IrO$_3$}
\newcommand{\lio}{Li$_2$IrO$_3$}
\newcommand{\aio}{$A_2$IrO$_3$}
\newcommand{\hlio}{H$_3$LiIr$_2$O$_6$}
\newcommand{\rucl}{$\alpha$-RuCl$_3$}

\newcommand{\Hc}{H_{\rm c}}
\newcommand{\TN}{T_{\rm N}}

\newcommand{\nth}{n_{\rm th}}
\newcommand{\diag}[1]{\mathrm{diag}\left(#1\right)}
\newcommand{\lllangle}{\langle\!\langle\!\langle}
\newcommand{\rrrangle}{\rangle\!\rangle\!\rangle}
\newcommand{\iu}{\mathrm{i}}

\graphicspath{{./}{./plots/}}

\begin{document}

\topical{Heisenberg-Kitaev physics in magnetic fields}

\author{Lukas Janssen and Matthias Vojta}

\address{Institut f\"ur Theoretische Physik and W\"urzburg-Dresden Cluster of Excellence ct.qmat, Technische Universit\"at Dresden, 01062 Dresden, Germany}

\eads{\mailto{lukas.janssen@tu-dresden.de}, \mailto{matthias.vojta@tu-dresden.de}}

\begin{abstract}
Magnetic insulators in the regime of strong spin-orbit coupling exhibit intriguing behaviors in external magnetic fields, reflecting the frustrated nature of their effective interactions. We review the recent advances in understanding the field responses of materials that are described by models with strongly bond-dependent spin exchange interactions, such as Kitaev's celebrated honeycomb model and its extensions. We discuss the field-induced phases and the complex magnetization processes found in these theories and compare with experimental results in the layered Mott insulators \rucl\ and \nio, which are believed to realize this fascinating physics.
\end{abstract}


\submitto{\JPCM}

\ioptwocol

\maketitle

\mainmatter

\tableofcontents
\markboth{{Heisenberg-Kitaev physics in magnetic fields}}{}

\section{Introduction}	   \label{sec:introduction}

Research in quantum magnetism has developed into a number of fascinating directions, many of which involve new forms of order and disorder \cite{starykh13,savary_rop17,kanoda_rmp17}. An important ingredient is spin-orbit coupling: For a long time considered as a minor perturbation, it has become clear that spin-orbit coupling can generate entirely new states of matter. In the context of insulating magnets, key contributions have been a seminal theoretical paper of Kitaev \cite{kitaev2006}, who presented the exact solution to a two-dimensional model with exchange frustration which realizes a $\mathbb{Z}_2$ quantum spin liquid, and the proposal of Chaloupka, Jackeli, and Khalliulin \cite{jackeli2009, chaloupka2010} on possible materials realizations of the  bond-anisotropic interactions underlying the Kitaev model.

Subsequently, a number of materials, with $4d$ or $5d$ magnetic ions placed on a layered honeycomb lattice, have been synthesized and investigated by a variety of methods \cite{rau2016, trebst2017, winter2017b, hermanns2018}. Most notable are \rucl, \nio, and different polytypes of \lio: While these materials display magnetic order at lowest temperatures (with the exception of \hlio\ \cite{kitagawa2018} which, however, appears to be rather disordered \cite{yadav2018, li_valenti2018}), various experimental findings suggest that they are located in proximity to a genuine spin-liquid phase \cite{banerjee2016a, banerjee2016b, nasu2016, do2017}. In parallel, the minimal microscopic model relevant to describing these materials, known as Heisenberg-Kitaev model, has been studied in detail \cite{chaloupka2013, rau2014a, osorio2014, gohlke2017}.

Both experiment \cite{majumder2015, kubota2015, johnson2015, das_sebastian2019} and theory \cite{yadav2016, janssen2016, chern2017, winter2018} indicate non-trivial and highly anisotropic behavior in an applied magnetic field, originating from the interplay of magnetic frustration and spin-orbit coupling. For \rucl, it has been found that a moderate magnetic field applied parallel to the Ru honeycomb plane suppresses the magnetic order and induces a gapped quantum paramagnetic state \cite{leahy2017, baek2017, sears2017, wolter2017, zheng2017, jansa2018, hentrich2018}. In the vicinity of this field-induced transition, fascinating properties such as excitation continua in neutron scattering \cite{banerjee2018} and evidence for additional intermediate phases \cite{kelley2018b, kasahara2018b} have been reported. Perhaps most intriguingly, for tilted fields, evidence for an approximately half-quantized thermal Hall response indicative of a chiral edge mode of Majorana fermions has been found in an intermediate-field regime \cite{kasahara2018b}.
These findings, together with ab-initio calculations \cite{kim2015, kim2016, winter2016, yadav2016, wangdong2017}, have in turn been used to refine the microscopic modelling of \rucl \cite{chaloupka2016, ran2017, winter2017a}, and to theoretically study the various field-induced phases and transitions in the relevant microscopic models \cite{janssen2017}. Similar investigations, both experimentally \cite{ruiz2017, choi2019} and theoretically \cite{lee2014b, rousochatzakis2018}, have also been performed for \aio\ compounds, with $A = \mathrm{Na}, \mathrm{Li}$.
In addition, model variations, both in parameter space \cite{rousochatzakis2017, catuneanu2018, gohlke2018, gordon2019} and on other lattices \cite{kimchi2014a, cook2015, becker2015, rousochatzakis2016, shindou2016, mizoguchi2016}, have been proposed and studied, which have unraveled exciting phenomenology and opened a broader view on the physics of spin-orbit-coupled magnets.

Together, these developments have promoted Heisenberg-Kitaev physics to a central item in the field of frustrated magnetism. It is the purpose of this review article to summarize important insights, in particular concerning the influence of external magnetic fields and the associated field-induced phenomena. On the theory side, the focus will be on analytical and numerical results for spin models involving Kitaev interactions; on the experimental side, we mainly concentrate on \rucl\ for which most extensive data is available, but also mention results obtained for \nio.

\subsection{Outline}

The body of this article is organized as follows:
To set the stage, the behavior of simple antiferromagnets in applied magnetic fields is summarized in Sec.~\ref{sec:primer}, together with general considerations on the modifications induced by spin-orbit coupling.
Sec.~\ref{sec:theory} is devoted to theory: It introduces the Heisenberg-Kitaev model and its extensions, discusses their phase diagrams under the influence of magnetic fields, and outlines experimentally relevant properties such as anisotropic susceptibilities, magnetization curves, and excitation spectra. Special emphasis is put on novel field-induced phases.
The following Sec.~\ref{sec:materials} summarizes important experimental results on phases, phase transitions, and anisotropic responses for \rucl\ and \nio\ compounds and confronts them with theoretical predictions.
An outlook, pointing to open questions and possible directions for future research, closes the paper.
A detailed appendix covers spin-wave theory for the relevant models in an applied magnetic field, which forms the basis for a semiclassical analysis.

\section{Primer: Canted antiferromagnets and spin-orbit coupling}   \label{sec:primer}

To understand the rich phenomenology of Heisenberg-Kitaev systems in external magnetic fields, it is instructive to first review the effect of spin-orbit coupling on the in-field behavior of antiferromagnets.

\subsection{Unfrustrated Heisenberg antiferromagnets in a magnetic field}

\begin{figure}[t]
\includegraphics[width=\linewidth]{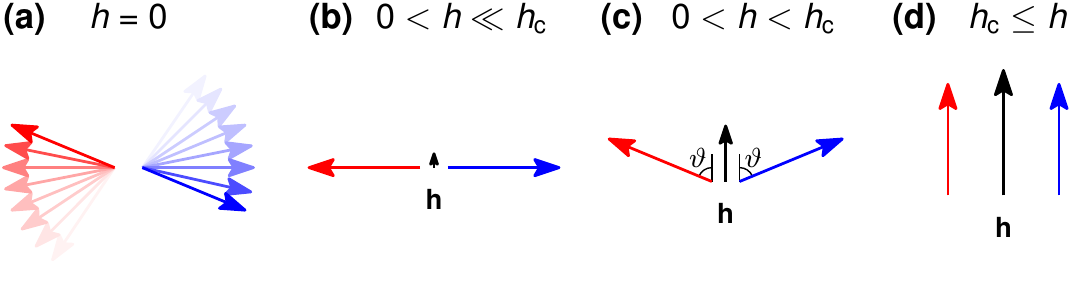}
\caption{Homogeneous canting of antiferromagnetic Heisenberg spins (red and blue arrows) in an external magnetic field $\mathbf h$ (black arrow). At $h = 0$ (a), there is an SU(2) degeneracy that is lifted in the presence of a finite external field, for which the spins align perpendicular to the field axis (b). At finite fields, the spins cant towards the field axis (c) and exhibit a continuous transition towards the high-field polarized phase at $h = h_\mathrm{c}$~(d).}
\label{fig:canting}
\end{figure}

As a minimal system, consider a nearest-neighbor Heisenberg model with SU(2) spin symmetry, described by the Hamiltonian
\begin{equation}
	\mathcal H_{\mathrm H} = J \sum_{\langle i j \rangle} \mathbf S_i \cdot \mathbf S_j,
\end{equation}
with an antiferromagnetic coupling $J > 0$ and the sum runs over nearest neighbors $\langle i j \rangle$ of some bipartite lattice. $\mathbf S_i$, $i=1,\dots,N$, represents a spin $1/2$ located at the \mbox{$i$-th} lattice site.
The response of such a Heisenberg antiferromagnet to a homogeneous external magnetic field $\mathbf H$ is well known: Assume a Zeeman coupling,
\begin{equation}
	\mathcal H_{\mathrm Z} = - \mathbf h \cdot \sum_{i=1}^N \mathbf S_i,
\end{equation}
with $\mathbf h \equiv \mu_\mathrm{B} \mu_0 g \mathbf H$, where $\mu_\mathrm{B}$ is the Bohr magneton and $g$ the symmetric (but potentially anisotropic) $g$ tensor.
An antiferromagnet has a vanishing total magnetization $\mathbf m \equiv \frac{1}{N} \sum_{i} \langle \mathbf S_i \rangle = 0$ at zero field, and there is an SU(2) degeneracy of ground states characterized by the direction of the staggered magnetization $\mathbf m_\mathrm{stagg} = \frac{1}{N} \sum_i (-1)^i \langle \mathbf S_i \rangle$. The staggered magnetization $\mathbf m_\mathrm{stagg}$ is the order parameter for the simple N\'eel antiferromagnet, which we assume to be the ground state of $\mathcal H_\mathrm{H}$.
An infinitesimal external magnetic field $\mathbf h$ partially lifts the degeneracy and selects the state that is characterized by the largest susceptibility. For Heisenberg systems, this means that the spins will align perpendicular to the field, $\mathbf m_\mathrm{stagg} \perp \mathbf h$.
For finite fields, the spins will cant towards the magnetic field axis with a homogeneous canting angle $\vartheta(h) \equiv \vartheta_i(h) = \angle (\langle\mathbf S_i\rangle, \mathbf h)$, with $\vartheta(h=0) = \pi/2$ and monotonously decreasing for $h>0$.
At some critical field strength $h = h_\mathrm{c}$, there will be a continuous transition towards the high-field polarized state, in which all spins are aligned along the magnetic-field axis, $\langle \mathbf S_i \rangle \parallel \mathbf h$ and $\vartheta(h \geq h_\mathrm{c}) = 0$. The canting of Heisenberg spins is illustrated in Fig.~\ref{fig:canting}.

\subsection{Spin-orbit couplings and magnetic field: General considerations}

\begin{figure}[t]
\includegraphics[width=\linewidth]{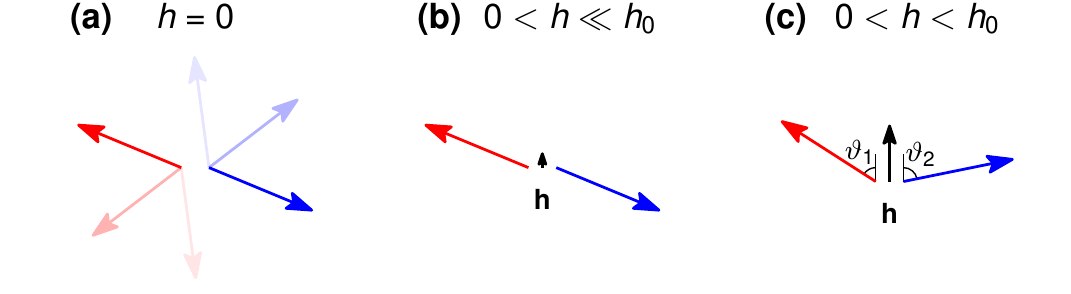}
\caption{Inhomogeneous canting of antiferromagnetic spins in the presence of spin-orbit coupling. At zero field (a), there is only a discrete set of ground states due to the absence of a continuous spin rotation symmetry. A perpendicular alignment of the spins to a small external field is therefore generically not possible (b). At finite fields, inequivalent spins will have different canting angles (c). At some higher field $h_0$, there will a transition towards another ordered or a disordered state (not shown).}
\label{fig:canting2}
\end{figure}

In the presence of spin-orbit coupling, however, the SU(2) spin rotation symmetry is typically broken down to combined discrete spin-lattice rotations. For instance, on the honeycomb lattice, Heisenberg-Kitaev systems exhibit at most only a residual $C_3^*$ symmetry of $2\pi/3$ spin rotation about the $[111]$ axis in spin space combined with a $2\pi/3$ lattice rotation about one site \cite{jiang2011, janssen2017}.
For systems with strong spin-orbit coupling, the response to an external magnetic field will therefore strongly depend on the field direction.
For instance, a spin alignment perpendicular to a small applied field, as in SU(2)-symmetric systems, will only be possible for selected field directions that happen to be perpendicular to the zero-field ordered moments. For generic field directions, inequivalent spins will have different canting angles $\vartheta_i$, with some of the $\vartheta_i > \pi/2$ for small $h$.
For finite fields, such states will therefore cant inhomogeneously towards the magnetic-field axis, see Fig.~\ref{fig:canting2}.
For strongly frustrated systems, a number of different states are energetically (nearly) degenerate. The canted versions of the zero-field ordered states that do not allow a homogeneous canting will therefore compete at finite fields with other states (metastable at zero field), for which a homogeneous canting is possible. This leads to interesting quantum phase transitions and nontrivial field-induced intermediate phases.

In the high-field phase in which all spins are parallel,%
\footnote{In this review, we define the ``high-field phase'' as the phase which preserves all symmetries and is adiabatically connected to the fully polarized state occurring at $h\to\infty$ (in which case $\langle \mathbf S_i \rangle \parallel \mathbf h$). In the literature, this phase is sometimes also referred to as ``forced-ferromagnetic'' or ``(partially) polarized" phase.}
$\langle \mathbf S_i\rangle \equiv \mathbf m$, the single-magnon state will no longer be an eigenstate of the Hamiltonian when SU(2)-breaking terms are present. In spin-orbit-coupled systems, quantum fluctuations are therefore present (and potentially sizable) at all finite fields $h < \infty$, and the magnetization will only be fully saturated in the strict limit $h \to \infty$.
We also note that in the presence of certain SU(2)-breaking interactions, the magnetization $\mathbf m$ in the high-field phase is no longer parallel to the field itself, as discussed below (see Sec.~\ref{subsec:kitaev-gamma}).

In the following, we will argue that Heisenberg-Kitaev systems offer a promising opportunity to investigate this rich and novel physics.

\section{Theory: Heisenberg-Kitaev models in external fields}   \label{sec:theory}

\subsection{Honeycomb-lattice Kitaev model}
\label{sec:th_kitaev}

\begin{figure}[tb]
\includegraphics[width=\linewidth]{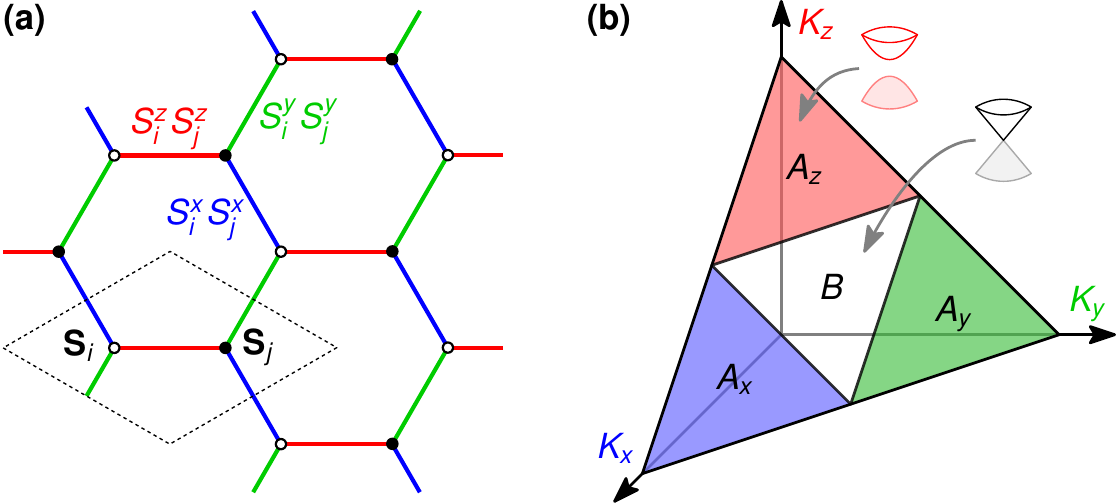}
\caption{
(a) Three inequivalent nearest-neighbor bonds on the honeycomb lattice, labeled as $\gamma = x, y, z$. On a $\gamma$ bond, the Kitaev spin-spin exchange interaction is $\mathcal H_{ij}^\gamma = K_\gamma S^\gamma_i S^\gamma_j$. The dashed parallelogram denotes the unit cell.
(b) Phase diagram of the Kitaev model on the plane $K_x + K_y + K_z = \mathrm{const}$. The $A$ phases are gapped, while the $B$ phase has gapless Majorana excitations at zero external field.
}
\label{fig:kitaev}
\end{figure}

The celebrated honeycomb-lattice Kitaev model describes a system of spins $1/2$ subject to a nearest-neighbor Ising-type interaction, with the quantization axis depending on the bond direction \cite{kitaev2006}. The Hamiltonian can be written as%
\footnote{Some theory works (including Ref.~\cite{kitaev2006}) define the Kitaev Hamiltonian in terms of Pauli spin matrices $\sigma^\gamma_i$, which are related to our dimensionless spin-1/2 operators as $S^\gamma_i \equiv \frac{1}{2} \sigma_i^\gamma$.}
\begin{equation}
	\mathcal H_\mathrm{K} = \sum_{\gamma=x,y,z} K_\gamma \sum_{\langle i j\rangle_\gamma} S^\gamma_i S^\gamma_j.
\end{equation}
Here, $\gamma = x,y,z$ labels the three inequivalent nearest-neighbor bonds illustrated in Fig.~\ref{fig:kitaev}(a), with $K_x$, $K_y$, and $K_z$ denoting the respective coupling constants on these bonds.
The quantum ground state of this model is a spin liquid characterized by low-energy Majorana excitations (``spinons'') that are coupled to a static $\mathbb Z_2$ gauge field \cite{kitaev2006}. While the gauge-field excitations (``visons'') are gapped, the Majorana spectrum may be either gapless or gapped, depending on the relative size of the Kitaev couplings $K_x$, $K_y$, and $K_z$. In the vicinity of the $C_3^*$-symmetric point $K_x = K_y = K_z$, the band gap closes with a linear dispersion at isolated points in the Brillouin zone, corresponding to the $B$ phase in Fig.~\ref{fig:kitaev}(b). 
Right at $K_x = K_y = K_z \equiv K$, the gap closing occurs at the high-symmetry $\mathbf K$ and $\mathbf K' = - \mathbf K$ points, i.e., at the corners of the hexagonal Brillouin zone, and the gap of a single flux (vison) is $\Delta_v \approx 0.038 |K|$ \cite{kitaev2006}.%
\footnote{Note that a spin flip creates a pair of fluxes on adjacent plaquettes. Consequently, the excitation spectrum shows features only above the two-flux gap $\Delta_{2v} \approx 0.067 |K|$ \cite{kitaev2006, knolle2014}.}
By contrast, the Majorana spectrum is fully gapped if the magnitude of one of the couplings is larger than the sum of the magnitudes of the other two, e.g., $|K_z| > |K_x| + |K_y|$, corresponding to the three different $A$ phases in Fig.~\ref{fig:kitaev}(b).
Turning to finite temperatures, the Kitaev model does not display a finite-temperature phase transition -- this is a general property of $\mathbb Z_2$ topological phases. As studied in detail in Ref.~\cite{nasu2015}, there are two thermal crossovers at temperatures $T^*$ and $T^{**}$, where the higher temperature $T^{**} \sim K$ can be associated with the fractionalization of spins into mobile Majorana fermions and $\mathbb Z_2$ fluxes, while the lower temperature $T^* \sim \Delta_v$ corresponds to the freezing of the gauge fluxes.

\begin{figure}[t]
\centering\includegraphics[width=.9\linewidth]{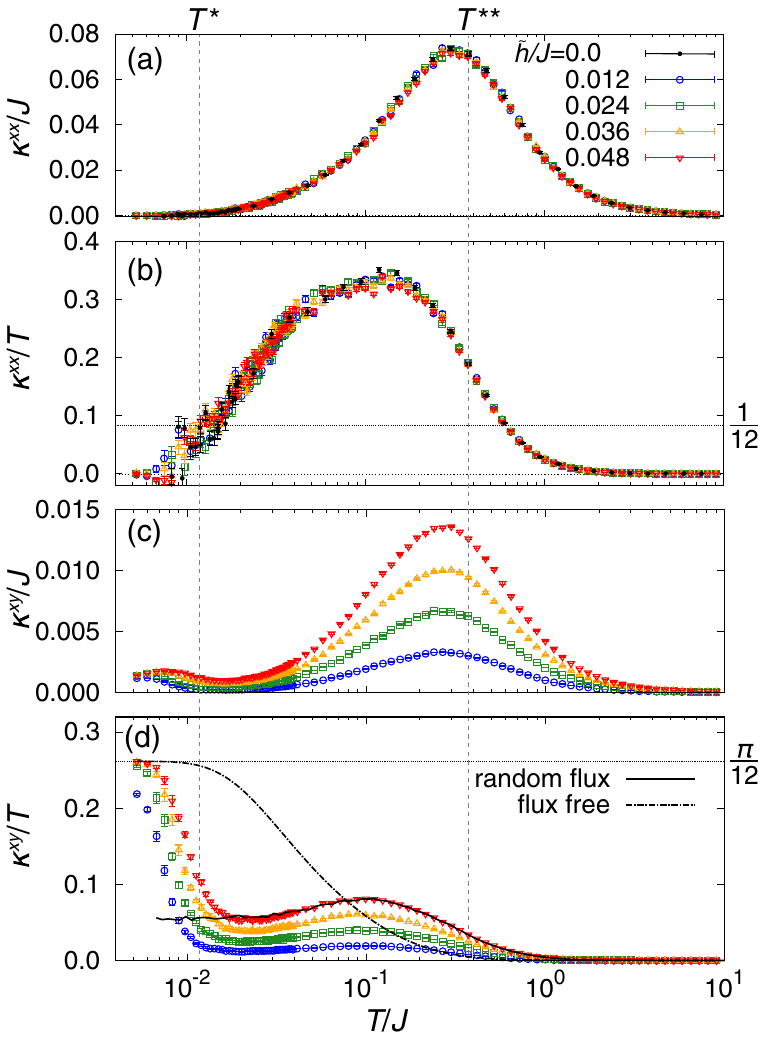}
\caption{Thermal transport coefficients in the Kitaev model in a small external field along the $[111]$ direction as function of temperature $T$ (in units of the Kitaev coupling denoted as $J$), from Quantum Monte Carlo simulations. Longitudinal conductivity, plotted as $\kappa_{xx}/J$ (a) and $\kappa_{xx}/T$ (b) in units of $k_\mathrm{B}^2/\hbar$. Transverse conductivity, plotted as $\kappa_{xy}/J$ (c) and $\kappa_{xy}/T$ (d).  Both the longitudinal and the transverse conductivities show crossover behaviors at the temperature scales $T^*$ and $T^{**}$. The higher scale $T^{**}$ can be associated with the fractionalization of spins into mobile Majorana fermions and $\mathbb Z_2$ fluxes. Below $T^{*}$, the fluxes become static. At the lowest temperatures, the transverse conductivity $k_{xy}/T$ saturates at the half-integer quantized value $\pi/12$, characteristic for Majorana edge states.
Reprinted from~\cite{nasu2017}.}
\label{fig:kitaev-transport-low}
\end{figure}

In the presence of an external magnetic field~$\mathbf h$, phase $B$ acquires a gap as well. For small fields, the Majorana spectrum can be computed perturbatively~\cite{kitaev2006}, leading to the dispersion
\begin{equation}
	\varepsilon(\mathbf q) \approx \sqrt{3 K^2 |\boldsymbol \delta \mathbf q|^2 + \Delta_s} \quad \mathrm{with} \quad \Delta_s \propto \frac{h_x h_y h_z}{K^2},
\end{equation}
where we have assumed the symmetric case $K_x = K_y = K_z$ and $\boldsymbol \delta \mathbf q = \mathbf q - (\pm \mathbf K)$ denotes the deviation from the wavevectors at which the gap closes at zero field. This phase is particularly interesting, as its effective excitations are characterized by nonabelian anyonic statistics \cite{kitaev2006}. In this phase, a spectral Chern number can be defined, in analogy to the topological invariant in the integer quantum Hall effect \cite{thouless1982}. When the magnetic field gaps out the Majorana fermions, the spectral Chern number is finite, which means that gapless chiral edge modes arise at the boundaries of the sample \cite{halperin1982, hatsugai1993}.
In contrast to the electronic Hall effect, however, in the present spin model, the low-energy Majorana excitations are charge-neutral, leaving only a conserved energy current along the edge. As the number of complex-fermion degrees of freedom is only half of the number of Majorana fermions, this leads to a half-integer \emph{thermal} Hall effect, with a quantized Hall conductivity $\kappa_{xy}/T = \frac{1}{2}[(\pi k_\mathrm{B}^2)/(6\hbar)]$ \cite{kitaev2006} in the low-temperature limit. This may be viewed as a unique signature of chiral Majorana edge modes.
At finite temperatures, the thermal Hall conductivity shows a crossover behavior \cite{nasu2017} at the two characteristic temperature scales $T^*$ and $T^{**}$, see Fig.~\ref{fig:kitaev-transport-low}.
We note that small perturbations away from the pure Kitaev limit change the cubic scaling of the gap $\Delta_s \propto |\mathbf h|^3$ into a linear dependence $\Delta_s \propto |\mathbf h|$ \cite{song2016}.

\begin{figure}
\includegraphics[width=\linewidth]{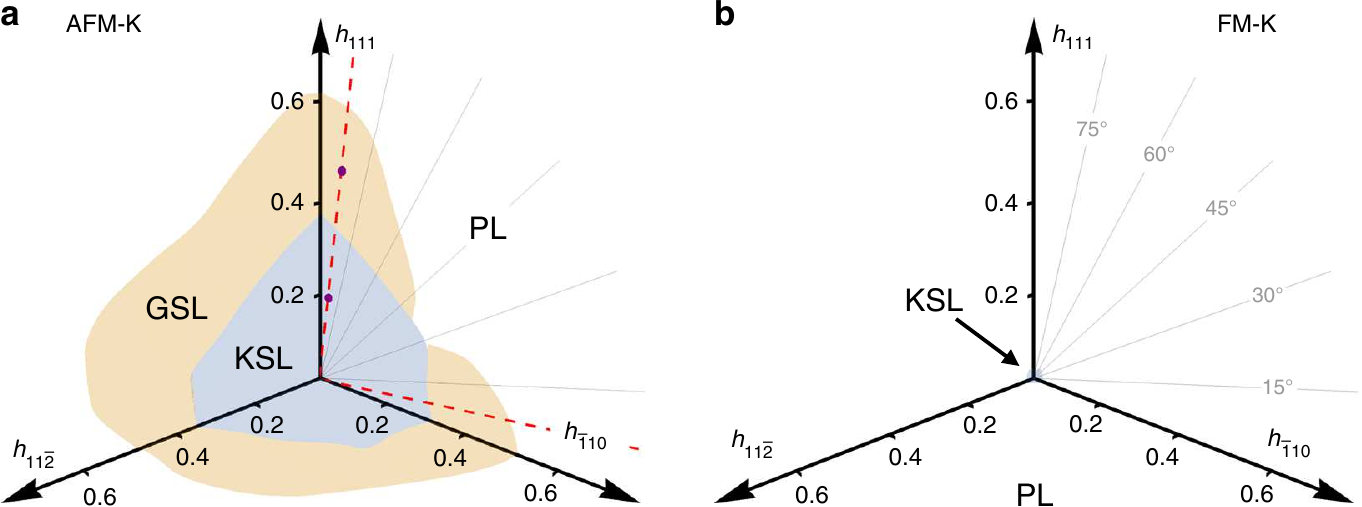}
\caption{Phase diagrams of $S=1/2$ Kitaev models in an external magnetic field for various field directions as obtained from exact diagonalization on a $24$-site cluster. The three axes correspond to the $[11\bar2]$, $[\bar 110]$, and $[111]$ directions in spin space. (a)~In the antiferromagnetic case, $K > 0$, a gapless spin liquid (GSL) appears between the gapped Kitaev spin liquid (KSL) and the high-field polarized (PL) phase. (b)~In the ferromagnetic case, $K < 0$, there is a direct transition at comparatively small fields from the KSL to the PL phase.
Reprinted from \cite{hickey2018}.}
\label{fig:kitaev-field}
\end{figure}

At elevated fields, perturbation theory is no longer applicable. At some finite $h$, one expects a transition to the high-field phase. While at zero field different signs of the Kitaev couplings are related to each other by a gauge transformation and therefore thermodynamically equivalent \cite{kitaev2006}, this does not apply to the field response. The semiclassical analysis shows that the antiferromagnetic Kitaev model with $K>0$ is much more stable towards the application of a field than the ferromagnetic Kitaev model with $K<0$ \cite{janssen2016}. In fact, an early numerical analysis of the $K<0$ case indicated a direct transition between the gapped version of the Kitaev spin liquid $B$ phase and the high-field phase at a small value of applied field, $h_{\rm c}/|K| \approx 0.02$ \cite{jiang2011}, cf.\ Fig.~\ref{fig:kitaev-field}(b).

\begin{figure*}
\includegraphics[width=\linewidth]{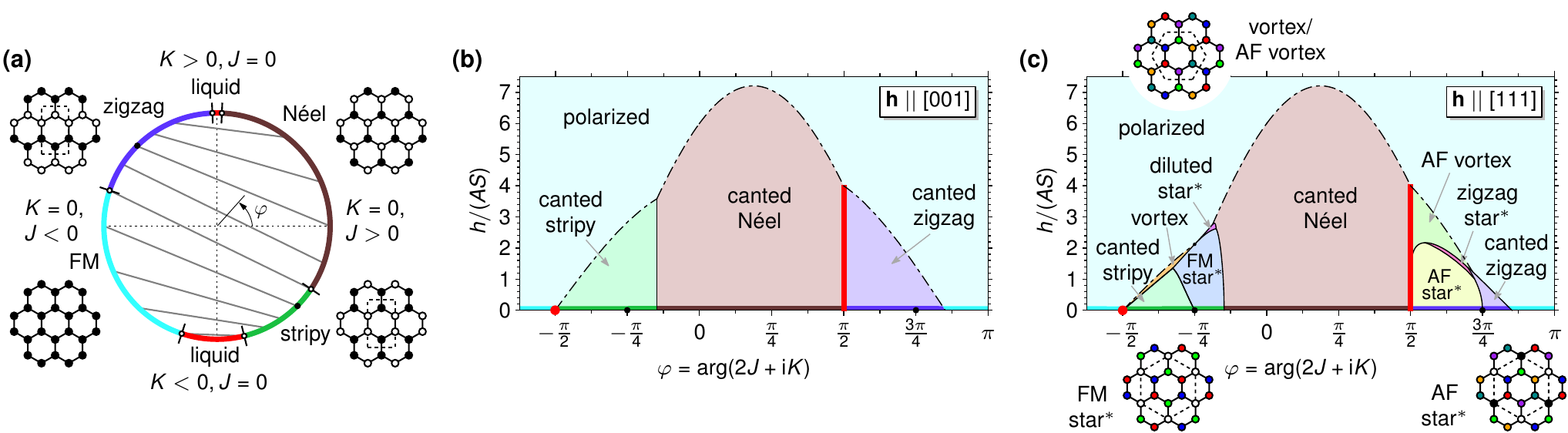}
\caption{(a) Phase diagram of the nearest-neighbor Heisenberg-Kitaev model for $S=1/2$ as obtained from exact diagonalization. The couplings have been parameterized as $J = A \cos \varphi$ and $K = 2 A \sin \varphi$. Besides the spin-liquid phases (red) near $\varphi = \pm \pi/2$, there are four magnetically ordered phases, the spin pattern of which are illustrated in the insets.
Gray lines connect points in the phase diagram that are related by a four-sublattice spin rotation.
Points with hidden SU(2) symmetry are marked with a black dot.
(b) Classical phase diagram ($S\to\infty$) of the Heisenberg-Kitaev model in an external magnetic field along one of the cubic axes, $\mathbf h \parallel [001]$. While the ordered phases survive in the classical limit, the Kitaev spin-liquid phases shrink to an isolated (red) line at $\varphi = \pi/2$ and a single (red) point at $\varphi = - \pi/2$.
(c) Same as (b) for $\mathbf h \parallel [111]$. Near the Kitaev limits, several new ordered phases are stabilized at intermediate field strength.
Phases marked with ``$*$'' denote multi-$\mathbf Q$ states.
Insets show spin patterns of three of the new phases, with the different colors representing inequivalent spins. The magnetic unit cells are shown dashed.
Panel (a) reprinted from \cite{chaloupka2013}; panels (b,c) adapted from \cite{janssen2016}.}
\label{fig:hk-field}
\end{figure*}

A number of recent works have focussed on the antiferromagnetic case $K>0$. Exact diagonalization and density matrix renormalization group studies~\cite{zhu2018, gohlke2018, ronquillo2018, hickey2018, jiang2018, zou2018, patel2018}, as well as parton mean-field approaches \cite{nasu2018, liang2018}, have provided mounting evidence for a new topological phase located between the low-field Kitaev spin liquid and the high-field phase, see Fig.~\ref{fig:kitaev-field}(a).
For a field along the $[111]$ direction, this intermediate phase roughly covers the field range $0.39<h/K<0.6$.
Although the extrapolation of the numerical data to the thermodynamic limit is challenging, the current consensus is that this intermediate phase is characterized by gapless spinons that interact via, potentially also gapless, gauge-field excitations. One possible scenario is that the intermediate phase realizes a U(1) spin liquid with a neutral Fermi surface \cite{jiang2018, zou2018, patel2018}. In this picture, the $\mathbb Z_2$ Kitaev spin liquid at small fields can be understood as a ``Higgsed-out'' version of the U(1) spin liquid, while the high-field phase can be obtained by monopole proliferation, leading to spinon confinement \cite{hickey2018}. The corresponding quantum phase transitions in this scenario are continuous and hence define exotic new quantum universality classes beyond the standard Landau-Ginzburg-Wilson paradigm.

\subsection{Nearest-neighbor Heisenberg-Kitaev model{$\!\!$}}

In real materials, further interactions in addition to the Kitaev coupling are present. These may stabilize other interesting ordered and disordered phases. The simplest extension of the pure Kitaev model takes the effects of a nearest-neighbor Heisenberg interaction into account:
\begin{equation}
	\mathcal H_\mathrm{HK} = J \sum_{\langle i j \rangle} \mathbf S_i \cdot \mathbf S_j + K \sum_{\gamma} \sum_{\langle i j \rangle_\gamma} S_i^\gamma S_j^\gamma,
\end{equation}
where we have again assumed $C_3^*$ symmetry with $K \equiv K_x = K_y = K_z$.

This model has been extensively studied at zero field by means of exact diagonalization \cite{chaloupka2010, chaloupka2013}, density matrix renormalization group \cite{jiang2011, gohlke2017}, projected entangled pair state simulations \cite{iregui2014}, and functional renormalization group \cite{reuther2011}. The zero-field phase diagram is illustrated as a function of the angle $\varphi = \arg(2J + \mathrm i K)$ in Fig.~\ref{fig:hk-field}(a).
In the vicinity of the Kitaev limits at $\varphi = \pi/2$ and $\varphi = 3\pi/2$, respectively, the Kitaev spin liquid remains stable upon the inclusion of a small Heisenberg coupling $|J| \ll |K|$.
For large Heisenberg coupling, $|J| \gg |K|$, the ground state has N\'eel antiferromagnetic order if $J>0$ and ferromagnetic order if $J<0$.
The phase diagram consists of two further ordered phases that are stabilized when $J$ and $K$ have similar magnitudes but opposite signs.
The zigzag ground state occurring for $J < 0$ and $K > 0$ is characterized by zigzag chains of parallel spins along $y$ and $z$ bonds, $z$ and $x$ bonds, or $x$ and $y$ bonds. In agreement with the $C_3^*$ symmetry, this state thus consists of three different domains, which we refer to as $x$, $y$, and $z$ zigzag.
In the stripy state for $J > 0$ and $K < 0$, parallel spins form stripes that are aligned perpendicular to either the $x$, $y$, or $z$ bonds of the honeycomb lattice, representing again three different domains.

The presence of these further ordered phases can be established in the vicinity of two distinguished so-called ``Klein'' points, marked by black dots in Fig.~\ref{fig:hk-field}(a).
By employing a four-sublattice spin rotation, it is possible to map the Heisenberg-Kitaev model at $\varphi = -\pi/4$ ($\varphi = 3\pi/4$) onto a dual Heisenberg model with $J < 0$ ($J>0$) \cite{khaliullin2005, chaloupka2010}.
For $\varphi = -\pi/4$, the exact ground state is therefore obtained by transforming the ferromagnetic ground state in the rotated basis back to the original spin basis. If, for instance, the ferromagnet in the rotated basis points along the $x$, $y$, or $z$ axis, the corresponding state in the original basis becomes the $x$, $y$, or $z$ stripy state \cite{chaloupka2010}. This also means that the stripy ground state is fluctuation-free at $\varphi = -\pi/4$, such that the corresponding staggered magnetization is fully saturated.
An analogous argument applies to the situation for $\varphi = 3\pi/4$, thereby establishing the zigzag ground state in the vicinity of this second Klein point. However, since the dual Heisenberg model is antiferromagnetic in this case, the corresponding staggered magnetization is significantly reduced to about 54\% of the classical value \cite{oitmaa1992, castro2006, farnell2014}.
The duality shows that the Heisenberg-Kitaev model at the two Klein points has a hidden SU(2) symmetry \cite{chaloupka2015} and the three domains of stripy and zigzag states, respectively, are consequently part of a continuous ground-state manifold, all members of which are energetically degenerate both in the classical and quantum cases. While the stripy and zigzag states are collinear single-$\mathbf Q$ states, this manifold includes also states that are characterized by noncollinear or noncoplanar spin patterns and/or multiple Bragg peaks in the first Brillouin zone (multi-$\mathbf Q$ states) \cite{janssen2016}.
Away from the special points of hidden SU(2) symmetry, the degeneracy is lifted. It has been shown that both quantum \cite{chaloupka2010, chaloupka2013} and thermal fluctuations \cite{price2012, price2013} generically stabilize the collinear stripy and zigzag states, respectively, to the detriment of the other, noncollinear and noncoplanar, states.

An external magnetic field, however, will not necessarily also favor canted version of collinear states. This will in particular be true for field directions that do not allow a homogeneous canting, i.e., when the direction of the zero-field ordered moments is not perpendicular to the field axis.
For the $x$, $y$, and $z$ stripy states in the present model, for instance, the spins are aligned along the cubic $x$, $y$, and $z$ axes. Consequently, these states will cant inhomogeneously towards a field if the latter is not perpendicular to at least one of the cubic axes. Another member of the ground-state manifold at $\varphi = -\pi/4$ is the ``FM star'' state, which is characterized by an 8-site magnetic unit cell, see inset in Fig.~\ref{fig:hk-field}(c). In this state, two spins each are aligned along the $[111]$, $[\bar 1 \bar 1 1]$, $[\bar 1 1 \bar 1]$, and $[1 \bar 1 \bar 1]$ axes.
In fact, one can show that this state has the largest susceptibility among the members of the hidden-SU(2) ground-state manifold, if the magnetic field is along the $[111]$ direction \cite{janssen2016}. This means that a $[111]$ field will drive a transition from the zero-field stripy state towards FM star if $\varphi$ is near $- \pi/4$. This applies both to the classical and the quantum limit, and a similar field-induced quantum phase transition should be expected for $\varphi$ near $3\pi/4$ in the regime of the zigzag zero-field ground state.
The full quantum phase diagram of the Heisenberg-Kitaev model in field has so far not been mapped out, but the classical analysis shows a variety of further interesting transitions and field-induced ordered phases for a field in the $[111]$ direction \cite{janssen2016, chern2017}, see Fig.~\ref{fig:hk-field}(c).
On the other hand, if the field is aligned perpendicular to one of the cubic axes, e.g., $\mathbf h \parallel [001]$ or $\mathbf h \parallel [\bar 1 1 0]$, then at least one domain of both stripy and zigzag states can cant homogeneously towards the magnetic field axes. In theses cases, no intermediate field-induced phases are to be expected, in agreement with the classical calculation, see Fig.~\ref{fig:hk-field}(b).

\subsection{Kitaev-Gamma model}
\label{subsec:kitaev-gamma}

Another interaction that is fully compatible with the $C_3^*$ symmetry of the Kitaev honeycomb model, and will therefore also generically be present in real materials, is the off-diagonal Gamma interaction~\cite{rau2014a}:
\begin{equation}
	\mathcal H_{\Gamma} = \Gamma \sum_{\gamma} \sum_{\langle i j \rangle_\gamma}
	\left(
	S_i^\alpha S_j^\beta + S_i^\beta S_j^\alpha
	\right),
\end{equation}
where $(\alpha, \beta, \gamma)$ is a permutation of $(x,y,z)$, such that $\alpha$ and $\beta$ label the two remaining directions on a $\gamma$ bond.

\begin{figure}
\includegraphics[width=\linewidth]{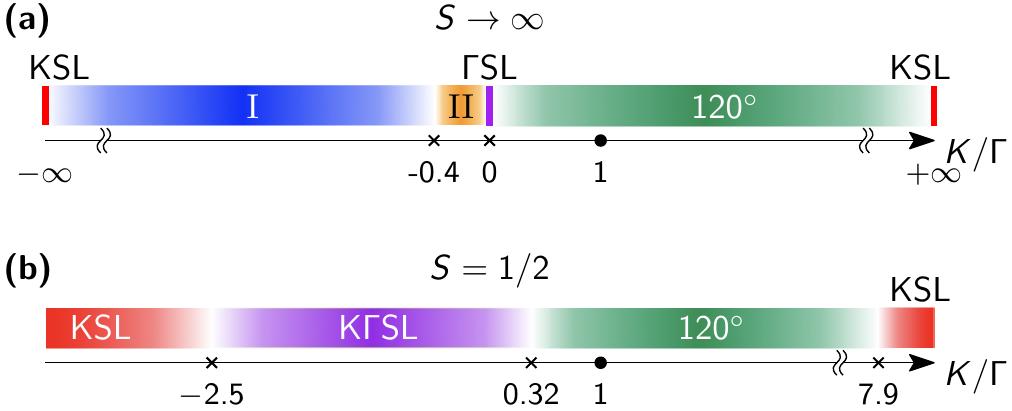}
\caption{Phase diagrams of the Kitaev-Gamma model for $\Gamma > 0$ (a) in the large-$S$ limit, as obtained from classical Monte-Carlo simulations \cite{rau2014b, janssen2017} and (b) for $S=1/2$, as obtained from 24-site exact diagonalization and density matrix renormalization group on a two-leg strip \cite{rau2014b, gordon2019}. For $K/\Gamma < 0$, the classical multi-$\mathbf Q$ and/or incommensurate phases ``I'' and ``II'' give way to the Kitaev spin liquid (KSL) and the Kitaev-Gamma spin liquid (K$\Gamma$SL). Points with hidden SU(2) symmetry are marked with black dots.}
\label{fig:kitaev-gamma}	
\end{figure}

In the classical limit, the Gamma-only model $\mathcal H_\Gamma$ has an extensive ground-state degeneracy, realizing a classical spin liquid \cite{rousochatzakis2017}.
Adding an infinitesimal perturbation to this Gamma spin liquid drives a transition towards a magnetically ordered state.
The classical Kitaev-Gamma model with Hamiltonian $\mathcal H_{\mathrm{K}\Gamma} = \mathcal H_\mathrm{K} + \mathcal H_\Gamma$ has a $120^\circ$ ground state for $K/\Gamma > 0$~\cite{rau2014a} and a multi-$\mathbf Q$ and/or incommensurate ground state for $K/\Gamma < 0$ \cite{rau2014b, janssen2017}.
The $S=1/2$ Kitaev-Gamma model has been studied using exact diagonalization~\cite{catuneanu2018} and density matrix renormalization group~\cite{gohlke2018, gordon2019} for $\Gamma > 0$. These works found an extended parameter region $-\infty < K/\Gamma < 0.32$ for which the model does not exhibit magnetic order at zero temperature. For $K/\Gamma < -2.5$, the disordered ground state appears to be adiabatically connected to the flux-free Kitaev spin-liquid state characterized by a unit expectation value of the plaquette operator, $\langle \hat W_p \rangle = 1$. Increasing $K/\Gamma$ towards the Gamma-only limit, the plaquette expectation value sharply drops below zero, $\langle \hat W_p \rangle \simeq -0.35$, suggesting a transition at $K/\Gamma \simeq -2.5$ towards a different quantum-disordered phase dubbed Kitaev-Gamma spin liquid \cite{kelley2018b, gordon2019}.
While the transfer matrix spectrum is suggestive of Majorana fermion excitations~\cite{gohlke2018}, the precise nature of this novel quantum spin liquid and its characteristic excitations is at present unknown.
In particular, it is not clear whether the state can be adiabatically connected to the Kitaev ground state or whether it realizes one of the various possible non-Kitaev spin liquids \cite{li2019}.
For antiferromagnetic Kitaev interaction with $0.32 < K/\Gamma < 7.9$, the quantum-disordered ground state gives way to the $120^\circ$ magnetically ordered state. 
Note that for $K/\Gamma = 1$ with $\Gamma > 0$, the model can be mapped by a six-sublattice transformation onto a ferromagnetic Heisenberg model~\cite{chaloupka2015}. The 120$^\circ$ state is therefore fluctuation-free at this point of hidden SU(2) symmetry.
Eventually, the antiferromagnetic Kitaev spin liquid is stabilized for $K/\Gamma > 7.9$.
The classical and quantum phase diagrams for the Kitaev-Gamma model with $\Gamma > 0$ are depicted in Fig.~\ref{fig:kitaev-gamma}.

In the pure Heisenberg-Kitaev model, the magnetic susceptibility $\chi_{\mathbf h}$, as well as the critical field $h_\mathrm{c}$ at which the transition to the high-field phase occurs, depends only weakly on the direction of the external field. However, when the off-diagonal Gamma interaction is included, a positive $\Gamma > 0$ naturally leads to a small magnetic susceptibility along the $[111]$ direction in the cubic basis (which in the materials corresponds to the direction perpendicular to the honeycomb layer, see Sec.~\ref{sec:directions}) and a large susceptibility along the $[11\bar 2]$ and $[\bar 1 10]$ directions (corresponding to two orthogonal in-plane directions)~\cite{janssen2017,gohlke2018}.
This behavior can be understood by realizing that $\Gamma > 0$ acts as an antiferromagnetic coupling for spins that are aligned along the out-of-plane $[111]$ direction, but as a ferromagnetic coupling for spins along the in-plane directions.
Consequently, in a model with sizable $\Gamma > 0$, the out-of-plane critical field $h_\mathrm{c}^{\perp}$ above which the high-field polarized state is stabilized is significantly larger than the in-plane critical field $h_\mathrm{c}^{\parallel}$.

Although the Gamma interaction respects the same crystal symmetries as the Heisenberg-Kitaev model, the latter possesses a further (accidental) discrete symmetry that is broken by a finite $\Gamma$. For certain field directions, this has been shown to lead to a finite transversal magnetization $\mathbf m_\perp \perp \mathbf h$ in the high-field phase, i.e., $\langle \mathbf S_i\rangle \equiv \langle \mathbf S \rangle$ is not parallel to $\mathbf h$ for all $h < \infty$~\cite{janssen2017}.

For intermediate fields $0 < h < h_\mathrm{c}$, due to the strong frustration and the broken SU(2) spin symmetry, one may expect metamagnetic transitions and novel field-induced phases for field directions that do not allow a homogeneous canting of the classically ordered ground states, as discussed in Sec.~\ref{sec:primer}. In the classical limit, this happens, for instance, if the field is aligned along the in-plane $[11\bar 2]$ direction~\cite{kelley2018b}.
For the $S=1/2$ case, a magnetic field with a significant component in the out-of-plane $[111]$ direction can also induce a transition towards a quantum-disordered phase that appears to be adiabatically connected to the Kitaev spin liquid \cite{gordon2019}. 
Further types of spin liquids have also been proposed for other field directions~\cite{liu2018}.

\subsection{Realistic models for \rucl\ and \nio}

The materials \rucl\ and \nio\ are spin-orbit-coupled Mott insulators that crystallize in a van-der-Waals layered structure with the magnetic ions Ru$^{3+}$ and Ir$^{3+}$, respectively, forming two-dimensional honeycomb lattices, for details see Sec.~\ref{sec:materials}.
The majority of the theoretical literature focusses on genuine two-dimensional models, assuming that the interlayer couplings can be neglected.
The generic exchange interaction $\mathcal H_{ij}$ between two sites $i$ and $j$ of the same honeycomb layer is constrained by symmetry~\cite{rau2014a, winter2016, rau2016, winter2017b}. Assuming local $C_{2h}$ symmetry of the $\gamma$ bond between (not necessarily nearest-neighbor) sites $i$ and $j$ leads to the form~\cite{winter2017b}
\begin{eqnarray}
	\mathcal H^{(\gamma)}_{ij} & = & J_{ij} \mathbf S_i \cdot \mathbf S_j + K_{ij} S_i^\gamma S_j^\gamma + \Gamma_{ij} \left( S_i^\alpha S_j^\beta + S_i^\beta S_j^\alpha \right)
	\nonumber \\ &&
	+ \Gamma_{ij}'\left(S_i^\gamma S_j^\alpha + S_i^\gamma S_j^\beta + S_i^\alpha S_j^\gamma + S_i^\beta S_j^\gamma \right),
\end{eqnarray}
with Heisenberg interactions $J_{ij}$, Kitaev interactions $K_{ij}$, and symmetric off-diagonal interactions $\Gamma_{ij}$ and $\Gamma_{ij}'$.
If trigonal distortions can be neglected, $\Gamma'_{ij}$ is expected to be small \cite{rau2014b}, and the coupling strengths $J_{ij}$, $K_{ij}$, and $\Gamma_{ij}$, respectively, are the same on bonds that are related by 120$^\circ$ lattice rotations. On the level of the nearest-neighbor interactions, this leaves us with three independent couplings $J$, $K$, and $\Gamma$, which are the same respectively on all three types of bonds $x$, $y$, and $z$.
This suggests the Heisenberg-Kitaev-Gamma (HK$\Gamma$) model with Hamiltonian
\begin{equation}
\mathcal H_\mathrm{HK\Gamma} = \mathcal H_\mathrm{HK} + \mathcal H_\Gamma + \dots
\end{equation}
as a minimal model in the maximally symmetric situation with undistorted honeycomb lattices. Here, the ellipsis denotes possible interactions beyond nearest neighbors.
The relative sizes of the interactions for \rucl\ and \nio\ have been estimated by using strong-coupling expansion~\cite{kim2015, kim2016, wangdong2017} and exact diagonalization~\cite{winter2016} of the relevant electronic Hamiltonian obtained from density functional theory, as well as by employing quantum chemistry computations~\cite{katukuri2014, yadav2016}.
Most of these works suggest a dominant ferromagnetic nearest-neighbor Kitaev interaction for both \rucl\ and \nio, supplemented with a sizable positive Gamma interaction for \rucl, and (potentially) a significant third-nearest-neighbor Heisenberg interaction that couples spins on opposite sites of a hexagon. We note that from these studies, no definite consensus has been reached concerning the detailed model parameters, as the results of ab-initio-based approaches sensitively depend on details of both the modelling and the assumed crystal structure; see \cite{janssen2017, winter2017b} for overviews.

Important insights into the relevant parameter regions for these interactions comes from the comparison with experimental results.
At low temperatures and zero field, \rucl\ and \nio\ have been found to display in-plane zigzag antiferromagnetic order \cite{choi2012, ye2012, sears2015, johnson2015}. There are three different basic mechanisms that are capable to realize this magnetic state: (1) For a strong antiferromagnetic Kitaev interaction, $K>0$, zigzag order can be stabilized by inclusion of a weak ferromagnetic Heisenberg interaction, $J<0$ \cite{chaloupka2013}; cf.\ Fig.~\ref{fig:hk-field}(a). (2) If the Kitaev interaction is ferromagnetic, $K<0$, a sizable antiferromagnetic third-nearest-neighbor Heisenberg interaction $J_3 > 0$ also induces zigzag order \cite{fouet2001, winter2016}. (3) Zigzag order can furthermore also be stabilized within a pure nearest-neighbor model for ferromagnetic Kitaev interaction if a sizable off-diagonal Gamma interaction $\Gamma > 0$ and/or $\Gamma' < 0$ is present \cite{rau2014a, rau2014b, ran2017, wangdong2017}.

\begin{figure}[t]
\centering \includegraphics[width=0.7\linewidth]{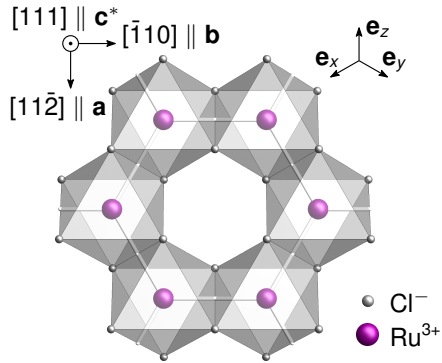}
\caption{Crystal structure of a \rucl\ layer with magnetic Ru$^{3+}$ ions sitting on the sites of a honeycomb lattice and nonmagnetic Cl$^-$ ions forming edge-shared octahedra. The $[111]$ axis in the cubic spin-space basis corresponds to the $\mathbf{c}^*$ axis, which is perpendicular to the honeycomb layer, while the $[\bar1 1 0]$ and $[11\bar2]$ axes correspond to the two in-plane directions parallel and perpendicular to a Ru-Ru bond. Reprinted from \cite{janssen2017}.}
\label{fig:rucl3}
\end{figure}

The response to an external magnetic field decisively depends on the particular zigzag stabilization mechanism, as well as on the respective field direction \cite{janssen2017, winter2018}.
In \rucl\ (\nio), the cubic basis vectors $\mathbf e_x$, $\mathbf e_y$, and $\mathbf e_z$, which represent the local spin quantization axes on the three different types of bonds in the Kitaev model, connect magnetic Ru$^{3+}$ (Ir$^{4+}$) ions with their nonmagnetic Cl$^-$ (O$^{2-}$) neighbors, the latter forming a network of edge-shared octahedra.
Without trigonal distortion, there are therefore three crystallographically inequivalent directions corresponding to the two in-plane directions $\mathbf a \parallel [1 1 \bar 2]$ and $\mathbf b \parallel [\bar 1 10]$ (perpendicular and parallel to Ru-Ru bonds, respectively) and the out-of-plane direction $\mathbf c^* \parallel [111]$; see Fig.~\ref{fig:rucl3}.
We note that the directions $\mathbf a$, $\mathbf b$, and $\mathbf c^*$ may or may not coincide with the crystallographic lattice vectors, depending on the actual three-dimensional crystal structure of the material; cf.\ Sec.\ \ref{sec:directions}.
For \rucl, the comparison of the theoretical results with the various experiments in field points to a HK$\Gamma$ model with dominant $K < 0$ and sizable $\Gamma > 0$ (Scenario 3), in qualitative agreement with the ab-initio results mentioned above~\cite{janssen2017}. A candidate parameter set, which has been successfully applied in different recent studies of \rucl\ is $(J, K, \Gamma, J_3) \simeq (-0.5,-5.0,+2.5,+0.5)$\,meV~\cite{winter2017a, wolter2017, winter2018}. \nio\ is believed to be characterized also by ferromagnetic Kitaev interactions, but a significantly smaller Gamma interaction (Scenario 2) \cite{janssen2017, das_sebastian2019}. We reiterate, however, that the precise values of the parameters, as well as the influence of other couplings and/or bond anisotropies, remain under debate to date.

\begin{figure}[t]
\centering\includegraphics[width=0.9\linewidth]{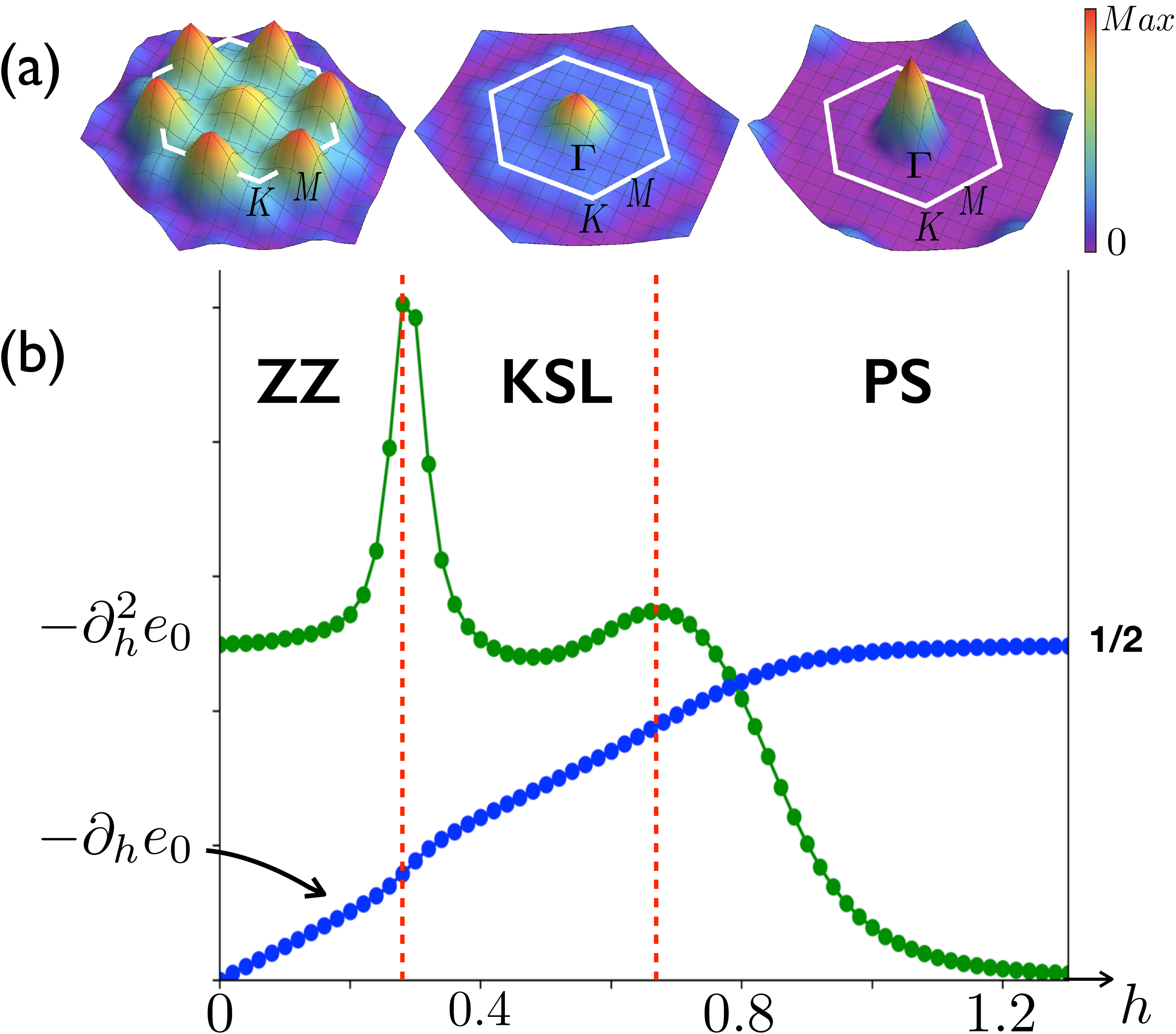}
\caption{Results for the $S=1/2$ HK$\Gamma$ model in an external magnetic field tilted $5^\circ$ away from the $[111]$ axis towards the $[11\bar2]$ axis. The model has a zigzag (ZZ) ground state for low fields $h \lesssim 0.3$ and stabilizes a disordered state that is adiabatically connected to the Kitaev spin liquid (KSL) at intermediate fields $0.3 \lesssim h \lesssim 0.7$, before the transition to the high-field polarized state (PS). (a) Static spin structure factor in the three different phases. (b) Magnetization (blue) and magnetic susceptibility (green) as function of $h$. Here, the parameters $(J,K,\Gamma) = (0,-0.938,0.347)$, supplemented by an additional $\Gamma'=-0.03$, have been used (all energies in units of $\sqrt{K^2 + \Gamma^2} \equiv 1$). Reprinted from \cite{gordon2019}.}
\label{fig:hkg-field}
\end{figure}

If frustration is large, the field may, for some field directions, induce transitions from the canted zigzag low-field ground state towards novel ordered or disordered intermediate phases, before the system eventually enters the high-field phase at high fields. This has been shown to occur generically for strong antiferromagnetic Kitaev interaction $K>0$ (Scenario~1) \cite{janssen2016, hickey2018, jiang2019} as well as for strong $\Gamma > 0$ (Scenario~3) \cite{yadav2016, kelley2018b}. In the latter case, a field with a significant out-of-plane component can stabilize a field-induced spin liquid that appears to be adiabatically connected to the Kitaev spin liquid \cite{gordon2019}; see Fig.~\ref{fig:hkg-field}.
This may be of relevance to the experiment of Ref.~\cite{kasahara2018b}, to be discussed in Sec.~\ref{subsec:rucl3-field}.

\begin{figure}[t]
\includegraphics[width=\linewidth]{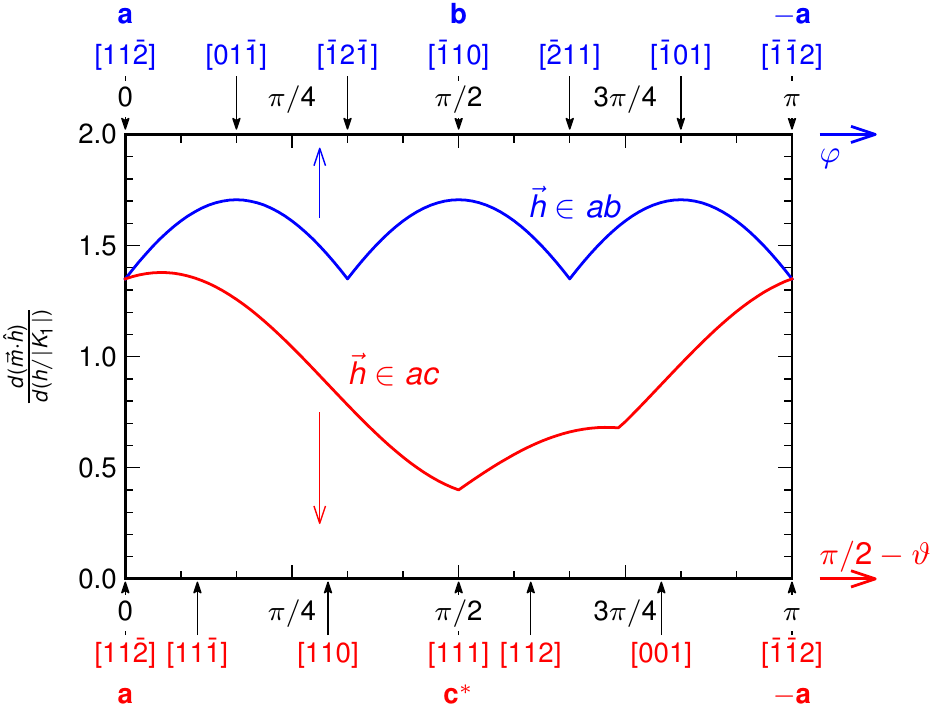}
\caption{Low-temperature susceptibility for a realistic model for \rucl\ as a function of field angle for small fields, neglecting a possible $g$ tensor anisotropy. For in-plane fields (blue), the susceptibility shows characteristic three-fold oscillations when rotating the field from $\mathbf a$ to $\mathbf {-a}$. When rotating from $\mathbf a$ to $\mathbf c^*$ (red), the susceptibility decreases significantly. Kinks arise from domain switching. Here, the parameters $(J,K,\Gamma) = (-0.5, -5.0, +2.5)\,\mathrm{meV}$, supplemented by an additional third-neighbor Heisenberg interaction $J_3 = 0.5\,\mathrm{meV}$ have been used.
Reprinted from \cite{janssen2017}. }
\label{fig:hkgamma-chi}
\end{figure}

In order to facilitate quantitative comparisons between the model calculations and the experimental data in external fields, knowledge of the $g$ tensor occurring in the definition of the effective field $\mathbf h \equiv \mu_\mathrm{B} \mu_0 g \mathbf H$ in terms of the physical external field $\mathbf H$ is indispensable. If trigonal distortions are weak, the $g$ tensor is expected to be isotropic, $g \approx \diag{g_{ab}, g_{ab}, g_{c^*}}$, with the in-plane component~$g_{ab}$ being of approximately the same size as the out-of plane component~$g_{c^*}$ \cite{agrestini2017}. If sizable trigonal distortions are present, it has been argued that the $g$ tensor is still diagonal in the $\{\mathbf a, \mathbf b, \mathbf c^*\}$ basis, but with anisotropic eigenvalues $g_{ab} \neq g_{c^*}$ \cite{chaloupka2016}. For \rucl, experiments~\cite{kubota2015, johnson2015, agrestini2017} and {\it ab initio} calculations~\cite{yadav2016} suggest $g_{ab}/g_{c^*} \gtrsim 1$, while for \nio\ it is expected that $g_{ab}/g_{c^*} \lesssim 1$ \cite{singh2010}. However, the debate on the actual size of the deviation from the isotropic case has not been settled to date. For \rucl, for instance, values for $g_{ab}/g_{c^*}$ ranging from $1.1$ \cite{agrestini2017} to $2.3$ \cite{yadav2016} have been reported.
We note that an estimate of the $g$ tensor from experimental data is complicated by the fact that a sizable off-diagonal Gamma interaction present in realistic models leads to a significant dependence of the magnetic response on the field direction, as discussed in Sec.~\ref{subsec:kitaev-gamma}. This is exemplified for a realistic model for \rucl\ in Fig.~\ref{fig:hkgamma-chi}, which shows the intrinsic dependence of the low-temperature magnetic susceptibility on the field angle.
Disentangling this intrinsic anisotropy, which arises from the bond-dependent interactions, from the extrinsic anisotropy that arises from an anisotropic $g$ tensor thus becomes rather model dependent.

\subsection{Topological magnons}

\begin{figure}[t]
\centering\includegraphics[width=0.85\linewidth]{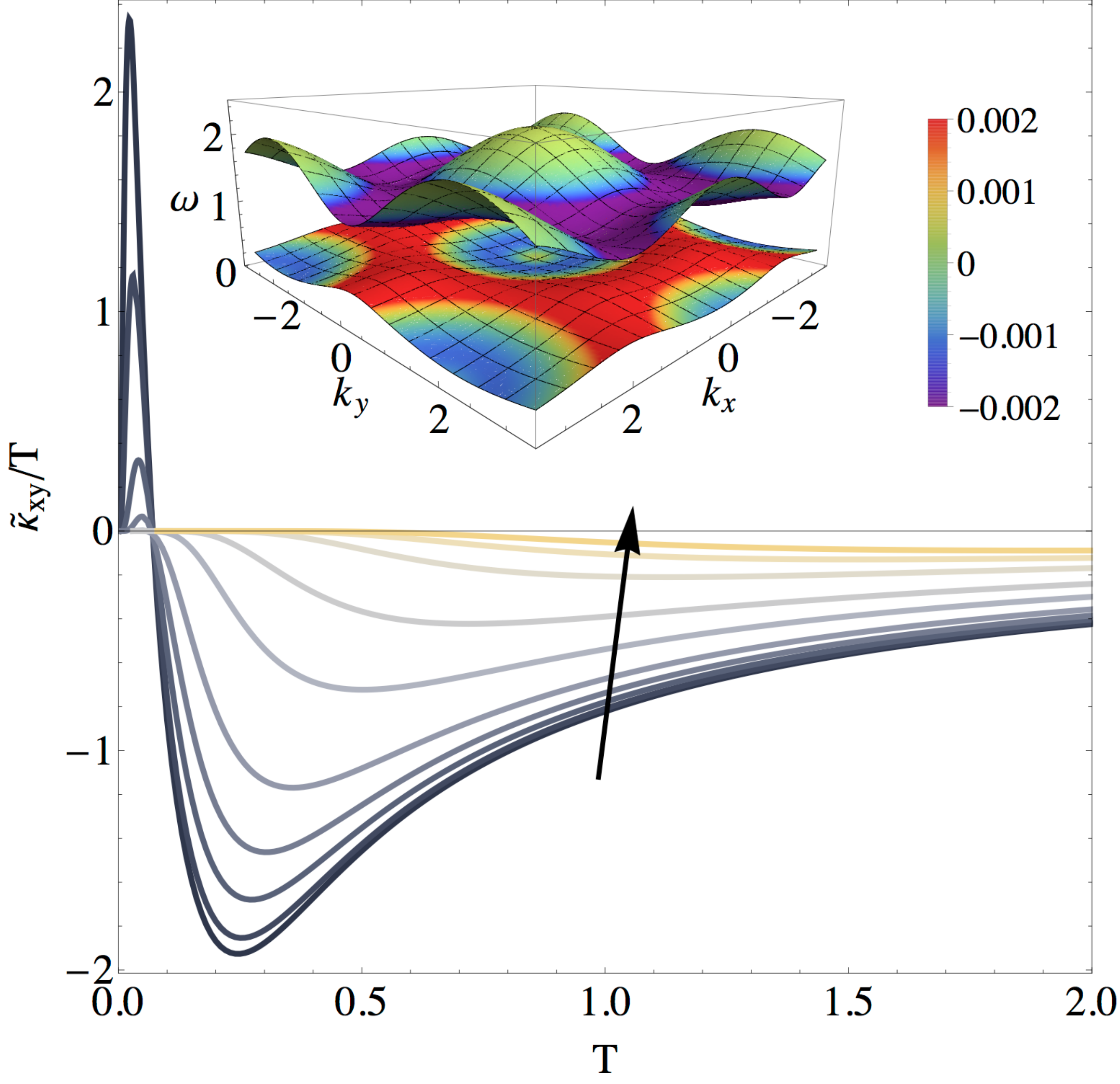}
\caption{Transverse thermal conductivity $\kappa_{xy}/T$ in units of $k_\mathrm{B}^2/\hbar$ in the high-field phase of the ferromagnetic Kitaev model in an external field along the $[111]$ direction, with $h/S = 0.01, 0.02, 0.05, 0.1, 0.2, 0.5, 1, 2, 3, 4$ (in the direction of the arrow). Inset: Dispersion of the two magnon bands at $h/S=0.1$, with the color indicating the Berry curvature. All energies in units of $|K| \equiv 1$.
Reprinted from~\cite{mcclarty2018}.}
\label{fig:kitaev-transport-high}
\end{figure}

At sizable fields above a certain critical field strength, the high-field polarized state is stabilized. While this phase may still be characterized by strong quantum fluctuations in the vicinity of the critical field as a consequence of the broken spin rotation invariance, quantum fluctuations will be suppressed in the infinite-field limit. This facilitates a controlled semiclassical description in terms of weakly interacting magnons.
As the magnetic unit cell in the high-field phase consists of two sites, the Brillouin zone features two magnon bands.
For a field in the $[111]$ direction, these two magnon bands in general do not touch each other, which allows to define a Berry curvature everywhere in the Brillouin zone~\cite{mcclarty2018, joshi2018}. Integration over the Berry curvature reveals nontrivial Chern numbers $\pm 1$ of the two bands, implying the existence of topologically protected chiral magnon edge modes that connect the two bulk bands.
If the field strength is not too close to the transition field, the magnon bands remain well-defined also upon taking magnon interactions into account, which suggests that their topological character may be of experimental relevance~\cite{mcclarty2018}. Figure~\ref{fig:kitaev-transport-high} displays the finite-temperature behavior of the thermal Hall response in the ferromagnetic Kitaev model, which shows a characteristic sign change resulting from the variation of the Berry curvature as a function of energy. However, in contrast to the situation in the spin-liquid phase at low fields shown in Fig.~\ref{fig:kitaev-transport-low}, the topological magnon edge states are gapped, leading to an exponential suppression of $\kappa_{xy}/T$ in the low-temperature limit.
If the spins are polarized in a direction away from the $[111]$ axis, such as $[001]$, the gap between the two magnon bands can be tuned to close and re-open as a function of the exchange interactions. This way, one can realize a topological transition across which the Chern numbers of the two bands change from $0$ to $\pm 1$ \cite{joshi2018}.
Similar topological transitions have recently also been predicted in the low-field zigzag phase~\cite{lu2018}.

\subsection{Other lattices and open questions}

Extended Heisenberg-Kitaev models can be defined on every lattice which has a coordination number that is divisible by three \cite{kimchi2014a}, and we list a few examples.

In the Heisenberg-Kitaev model on the triangular lattice, which may be relevant for the iridium perovskite Ba$_3$IrTi$_2$O$_9$, non-trivial spin textures emerge \cite{rousochatzakis2016, becker2015, catuneanu2015, shinjo2016, kos2017}. Their in-field behaviors remain largely unexplored to date.

On the three-dimensional hyperkagome lattice, bond-dependent Kitaev-like interactions have been proposed as a model for Na$_4$Ir$_3$O$_8$ \cite{chen2008, mizoguchi2016, shindou2016}.

The Heisenberg-Kitaev models on the three-dimensional hyperhoneycomb~\cite{lee2014a} and stripyhoneycomb lattices~\cite{kimchi2014b} exhibit ground-state phase diagrams that are analogous to those of the planar-honeycomb-lattice model, with three-dimensional analogs of zigzag and stripy magnetic orders. The lithium iridate polymorphs $\beta$- and $\gamma$-Li$_2$IrO$_3$ are spin-orbit-coupled magnetic insulators that crystallize in hyperhoneycomb and stripyhoneycomb geometries, respectively. These compounds, however, realize noncoplanar spiral orders \cite{biffin2014b, biffin2014c, modic2014, takayama2015} that are not captured by the pure Heisenberg-Kitaev models, similar to their planar-honeycomb-lattice polymorph $\alpha$-Li$_2$IrO$_3$ \cite{williams2016, choi2019}. 
These incommensurate orders can be realized within an anisotropic model of coupled zigzag chains \cite{kimchi2015}, or alternatively in a bond-isotropic HK$\Gamma$ model with equal magnitude of exchange couplings on all bonds~\cite{lee2015, lee2016}.

The incommensurate order in $\beta$-Li$_2$IrO$_3$ can be understood as a twisted version of a nearby commensurate order \cite{ducatman2018}. Application of a weak external magnetic field gradually reduces this order and simultaneously promotes a superimposed uniform zigzag order up until a characteristic field strength at which the incommensurate spiral order is completely suppressed, leaving behind only the field-induced canted zigzag order \cite{ruiz2017}.
The intensities of the Bragg peaks at the incommensurate and zigzag ordering wavevectors satisfy a sum rule, which may indicate that the two different states in fact represent intertwined components of the same order \cite{rousochatzakis2018}.

In \rucl, very recent inelastic neutron scattering experiments have determined the high-field magnon dispersion along the third, out-of-plane, direction, suggesting the interlayer spin exchange interactions to be of the order of (at most) a few percent of the in-plane couplings \cite{balz2019a}. A proper three-dimensional model that takes such interlayer couplings into account should reveal further interesting insights, in particular when the effects in external magnetic fields are considered.


\section{Materials: \rucl\ and \nio\ in external fields}
\label{sec:materials}


The purpose of this section is to discuss experimental results concerning Kitaev materials in applied magnetic fields, and to confront those with theory as described in the previous sections. We will restrict our attention to \rucl\ and \nio, for which extensive experimental data are available. We start with small-field properties, such as the linear-response susceptibility and the direction of ordered moments, and then discuss the behavior at elevated fields.


\subsection{Conventions for field directions}
\label{sec:directions}

At room temperature, \rucl\ \cite{johnson2015, cao2016} and \nio\ \cite{choi2012, ye2012} adopt a monoclinic crystal structure with space group $C2/m$. The low-temperature structure of \rucl\ has been a matter of some debate \cite{winter2017b}. Some of the most recent samples exhibit a structural phase transition in the region around 100--150 K \cite{kubota2015, glamazda2017, reschke2017, kelley2018a}, with the refinement of the neutron diffraction data indicating a rhombohedral structure with space group $R\bar 3$ at the lowest temperatures \cite{park2016}.

When specifying directions of fields and moments, we use, for both \rucl\ and \nio, the monoclinic notation as depicted in Fig.~\ref{fig:rucl3} for the example of \rucl.
We alert the reader that as a result of the confusion concerning the low-temperature structure of \rucl, different conventions have been used in the literature to label the directions of magnetic field and moments.
Most of the theory literature as well as Refs.~\cite{johnson2015} and \cite{cao2016} employ the monoclinic notation in which the two orthogonal in-plane axes $\mathbf a$ and $\mathbf b$ correspond to the directions perpendicular and parallel, respectively, to a nearest-neighbor Ru-Ru bond, as depicted in Fig.~\ref{fig:rucl3}.
Refs.~\cite{park2016,banerjee2018,kelley2018a,kelley2018b} instead use a trigonal notation in which the two in-plane basis vectors enclose an angle of 120$^\circ$ and are parallel to next-nearest-neighbor Ru-Ru bonds. The corresponding reciprocal lattice vectors hence enclose an angle of 60$^\circ$ and are both perpendicular to edges of the first Brillouin zone. The reciprocal $(1,\bar 2,0)$ direction in the trigonal notation [which is symmetry equivalent to $(2,\bar 1,0)$ and $(1,1,0)$] then corresponds to the $\mathbf a$ axis in the monoclinic notation of Fig.~\ref{fig:rucl3}, while the $(1,0,0)$ direction [which is symmetry equivalent to $(0,1,0)$ and $(\bar 1,1,0)$] corresponds to the $\mathbf b$ axis.
As third basis vector, we use the $\mathbf c^*$ axis, which corresponds to the out-of-plane direction perpendicular to $\mathbf a$ and $\mathbf b$ and agrees with $(0,0,1)$ in the trigonal notation. Note that in the monoclinic crystal structure, the actual third crystallographic axis~$\mathbf c$ lies in the $ac^*$ plane, but is tilted away from the $\mathbf c^*$ axis. E.g., in \nio\ at room temperature, the angle between $\mathbf c$ and $\mathbf a$ is $\beta = 109.0^\circ$ \cite{choi2012}, while $\mathbf c^* \perp \mathbf a$ per definition.

Finally, we note that the Kitaev model itself is formulated in terms of the cubic spin-space basis vectors $\mathbf e_x$, $\mathbf e_y$, and $\mathbf e_z$, which in \rucl\ (\nio) connect the central Ru (Ir) ions with the corners of their surrounding Cl (O) octahedra. The $[1 1 \bar 2]$ and $[\bar 1 1 0]$ directions in the spin-space basis correspond to the two in-plane axes $\mathbf a$ and $\mathbf b$ in the monoclinic notation, while the $[111]$ direction corresponds to the out-of-plane $\mathbf c^*$ axis; see Fig.~\ref{fig:rucl3}.

The different notations for three crystallographically inequivalent directions are summarized in Table~\ref{tab:directions}.

\begin{table}[t]
\caption{Crystallographically inequivalent directions in different notations. NN~bond: Nearest-neighbor Ru-Ru (Ir-Ir) bonds.
}\vspace{3pt}
\renewcommand{\arraystretch}{1.1}
\begin{tabular*}{\linewidth}{@{\extracolsep{\fill}}cccc}
\hline\hline
Monoclinic & Trigonal & Spin Space & \\
& (reciprocal) \\ \hline
$\mathbf a$ & $(1,\bar 2,0)$ & $[1 1 \bar 2]$ & \parbox[t]{6em}{in plane,\\ $\perp$~NN bond}\\
$\mathbf b$ & $(1,0,0)$ & $[\bar 1 1 0]$ & \parbox[t]{6em}{in plane,\\ $\parallel$~NN bond}\\
$\mathbf c^*$ & $(0,0,1)$ & $[1 1 1]$ & \parbox[t]{6em}{out of plane}\\
\hline \hline
\end{tabular*}
\label{tab:directions}
\end{table}


\subsection{Magnetic order and direction of moments}

The materials \rucl\ and \nio\ show magnetic order at low temperatures and zero field. In order to establish the responses of these materials to an external magnetic field, it is crucial to understand the types of order and the directions of the ordered moments.

\subsubsection[\rucl]{\rucl. }

\rucl\ displays zigzag antiferromagnetic order below $\TN=7$\,K \cite{sears2015, banerjee2016a}. In this single-$\mathbf Q$ state, the in-plane propagation wavevector equals one of the three $\mathbf M$ points in the Brillouin zone. In the out-of-plane direction, the magnetic order is fragile and easily altered by the formation of stacking faults. In high-quality single crystals, the magnetic refinement is consistent with a three-layer periodicity, dubbed ABC \cite{cao2016, park2016}. If the as-grown single crystals are mechanically deformed, a broader second transition at $\TN'=14$\,K is induced. Powder sample show only a single anomaly at 14\,K \cite{banerjee2016a}. This 14\,K transition is characterized by the formation of a two-layer periodicity (``AB'') \cite{johnson2015, banerjee2016a} and it has been associated with strong stacking disorder \cite{sears2015,cao2016}.

The direction of the ordered moments has been determined by polarized neutron diffraction: It lies in the $\mathbf{ac}^\ast$ plane, being tilted away from the $\mathbf{a}$ direction by an angle $\alpha_m \simeq 20^\circ$ \cite{balz2019}.
This is in reasonable agreement with the result deduced from the HK$\Gamma$ model if one assumes $\Gamma/K \simeq - 0.5$ and a $g$ tensor anisotropy of $g_{ab}/g_{c^*} \simeq 1.8$ \cite{kelley2018b}.

\subsubsection[\nio]{\nio. }

\begin{figure}[t]
\centering \includegraphics[width=0.9\linewidth]{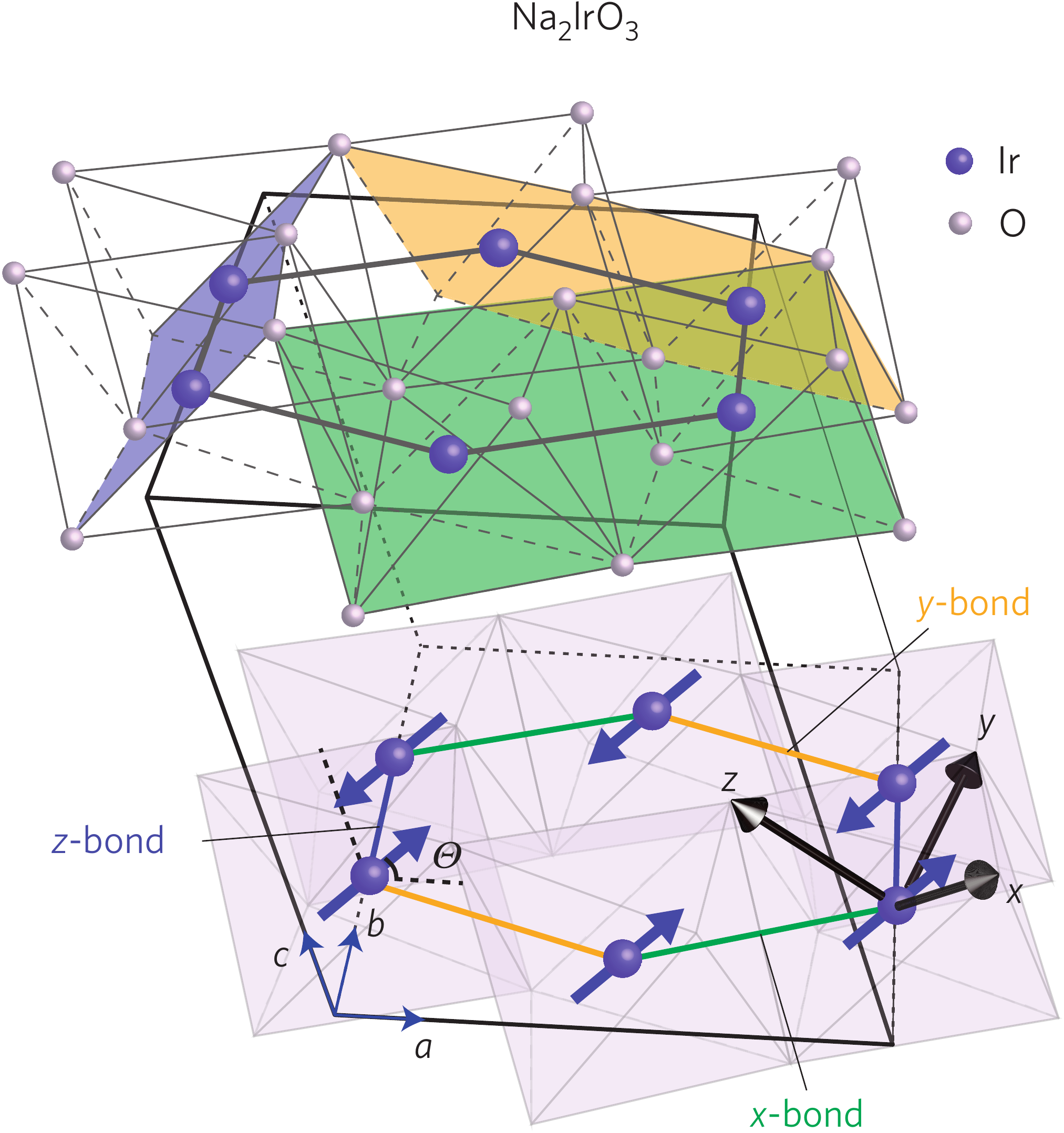}
\caption{Direction of spins in \nio. In the $z$ zigzag domain with antiferromagnetic $z$ bonds, the spins (blue arrows) lie in the $ac$ plane in the monoclinic basis, with an out-of-plane angle $\Theta$ that turns out to be such that the spins are close to the $[110]$ axis in the spin-space basis.
Reprinted from \cite{chun2015}.}
\label{fig:na2iro3-moments}
\end{figure}

\begin{figure*}[t]
\centering \includegraphics[width=\linewidth]{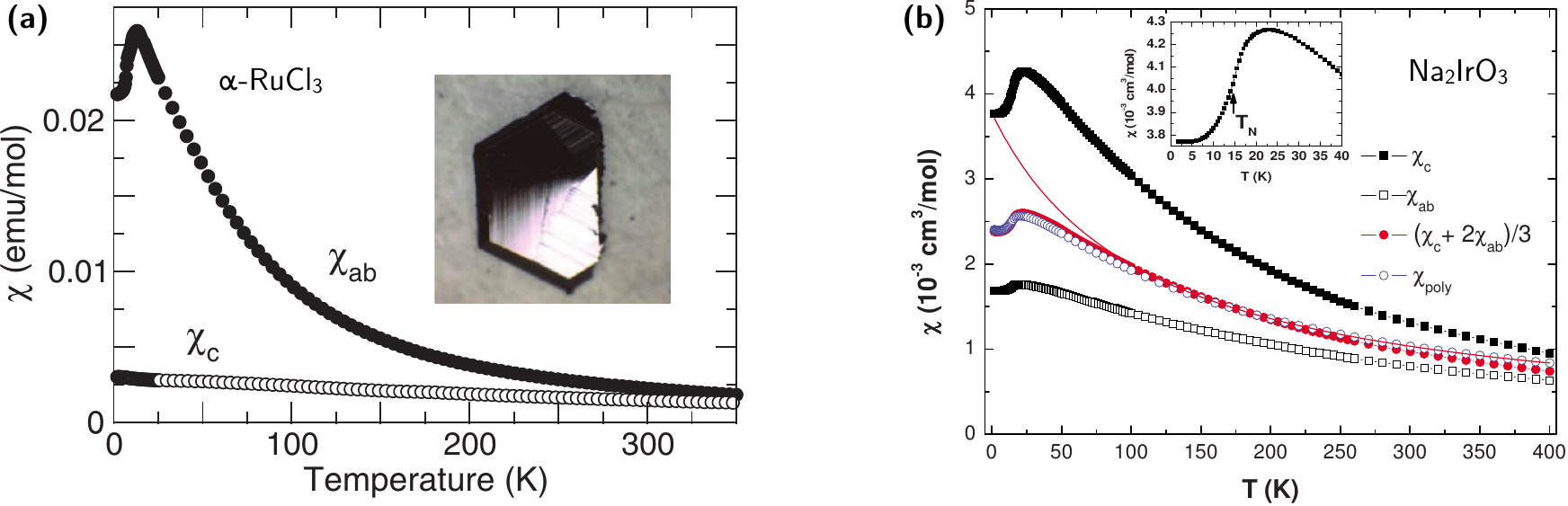}
\caption{Low-field susceptibilities as function of temperature in (a) \rucl\ (reprinted from \cite{sears2015}) and (b) \nio\ (reprinted from \cite{singh2010}). In \rucl, the out-of-plane susceptibility $\chi_{c}$ is much smaller than the in-plane susceptibility $\chi_{ab}$, while in \nio\ $\chi_{c} > \chi_{ab}$.}
\label{fig:rucl3-na2iro3-chi-T}
\end{figure*}

The known samples of \nio\ exhibit a single transition towards in-plane zigzag antiferromagnetic order at $\TN = 15$\,K \cite{singh2010, liu2011, singh2012, choi2012}.
In the out-of-plane direction, the neutron-diffraction data on single crystals are consistent with an antiferromagnetic AB stacking \cite{ye2012}.
The ordered moments again lie in the $\mathbf{ac}^*$ plane, but with a larger out-of-plane component as compared to \rucl. Using the resonant X-ray scattering data of Ref.~\cite{chun2015}, the tilting angle of the ordered moments $\mathbf m_i \propto g \mathbf S_i$ away from the in-plane $\mathbf a$ axis has been estimated as $\alpha_m \simeq 50^\circ$ \cite{chaloupka2016}. This result suggest a significantly smaller off-diagonal Gamma interaction in \nio\ as compared to \rucl\ \cite{sizyuk2016, janssen2017}.
The direction of the spins in \nio\ is illustrated in Fig.~\ref{fig:na2iro3-moments}.

\subsection{Magnetic anisotropy}

\begin{figure}[t]
\centering \includegraphics[width=0.9\linewidth]{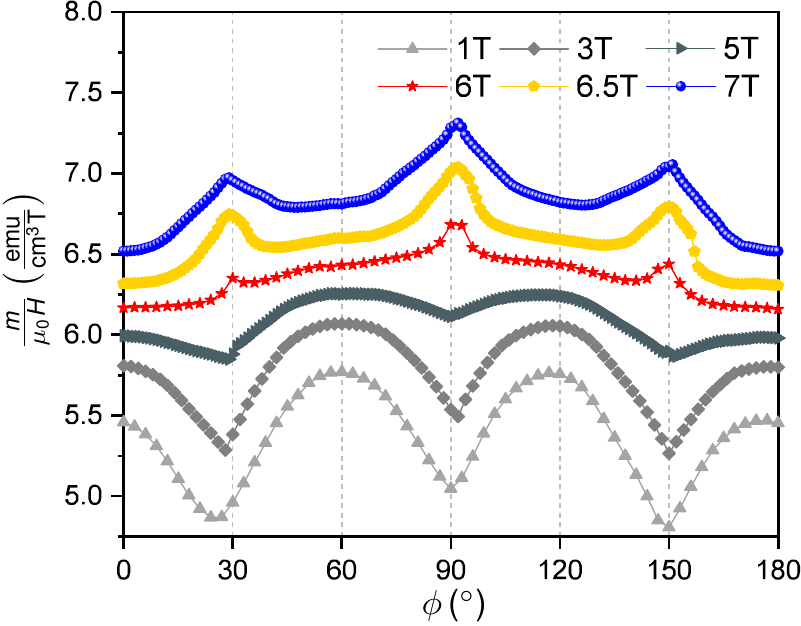}
\caption{Low-temperature susceptibility $m/(\mu_0 H)$ in \rucl\ as a function of the in-plane field angle $\phi$ for different field strengths. Here, $\phi = 0^\circ$ ($\phi = 90^\circ$) corresponds to a field $\mathbf H$ in the in-plane $\mathbf b$ ($-\mathbf a$) axis. The low-field curve should be compared with the theoretical result for the HK$\Gamma$ model (blue curve in Fig.~\ref{fig:hkgamma-chi}). Reprinted from \cite{kelley2018b}.}
\label{fig:rucl3-chi-phi}
\end{figure}

As explained in Sec.~\ref{subsec:kitaev-gamma}, bond-dependent interactions naturally lead to a strongly anisotropic magnetic response.

\subsubsection[\rucl]{\rucl. }

For \rucl, detailed measurements of the magnetic susceptibility at high temperature have been reported in Ref.~\cite{kelley2018a}; corresponding results at low temperatures are presented in Refs.~\cite{kelley2018b}. Earlier measurements have been communicated in Refs.~\cite{sears2015, johnson2015, majumder2015, kubota2015, banerjee2016a, baek2017, wolter2017}.

The most striking property is that the out-of-plane susceptibility $\chi_{c}$ for fields along $\mathbf c^*$ is significantly smaller than the in-plane susceptibility $\chi_{ab}$, see Fig.~\ref{fig:rucl3-na2iro3-chi-T}(a): Their ratio is slightly less than 2 at room temperature, grows to a factor of up to 10 close to the N\'eel temperature, and levels at around 5--8 deep in the zigzag ordered phase, depending on the particular in-plane direction \cite{kelley2018a}. The significant low-temperature anisotropy is unlikely to be an effect of an anisotropic $g$ tensor alone, which is expected to have an anisotropy ratio $g_{ab}/g_{c^*}$ of at most about~$2$ \cite{agrestini2017, yadav2016}, and therefore must be rooted in the interactions.
A natural explanation is provided by a positive off-diagonal Gamma term, which acts as an antiferromagnetic coupling for out-of-plane components of the spins, but as a ferromagnetic coupling for in-plane spin components \cite{janssen2017}. In the HK$\Gamma$ model with $\Gamma/K \simeq -0.5$, this intrinsic anisotropy leads to an additional factor of about 3--4 in the low-temperature susceptibility, which, together with a moderate $g$-tensor anisotropy, is consistent with the measured values in \rucl.

\begin{figure*}[t]
\includegraphics[width=0.9\columnwidth]{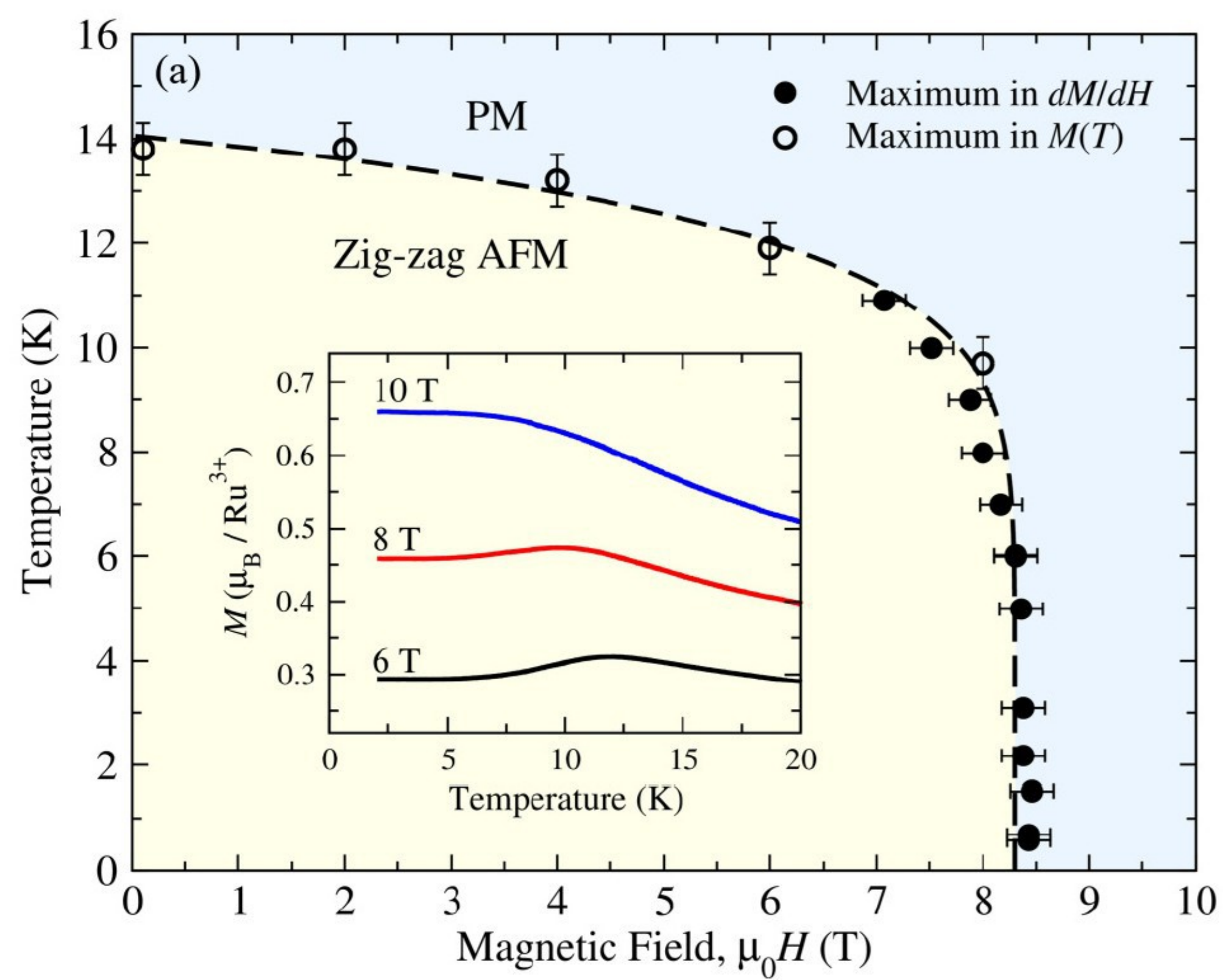}\hfill
\includegraphics[width=1.02\columnwidth]{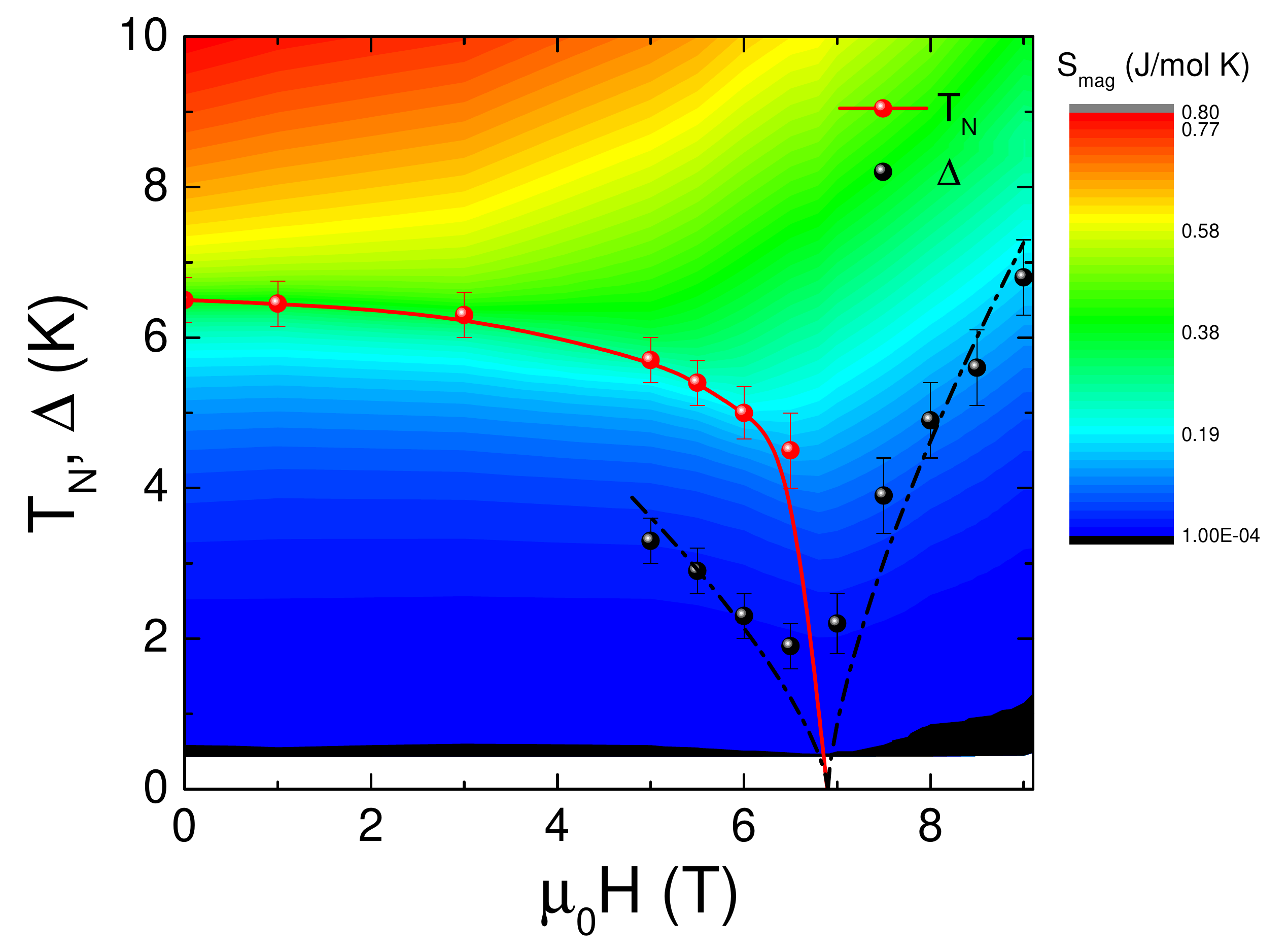}
\caption{
Phase diagrams of \rucl\ in in-plane magnetic fields
(a) extracted from magnetization measurements down to 0.7\,K (reprinted from \cite{johnson2015}) and
(b) extracted from specific-heat measurements down to 0.4\,K, with color-coded magnetic entropy  (reprinted from \cite{wolter2017}). The dashed lines indicate the extracted thermodynamic excitation gap.
}
\label{fig:pd_rucl}
\end{figure*}

Similarly, the critical field $H_\mathrm{c}$ at which the quantum transition towards the high-field phase occurs strongly depends on the field axis. It is significantly smaller for in-plane fields than for out-of-plane fields, in qualitative agreement with the low-field susceptibility \cite{johnson2015}. The dependence of $H_\mathrm{c}$ as a function of the out-of-plane field angle has recently been mapped out in detail \cite{modic2019}. It has a sharp maximum of about 35\,T occurring at an angle of around $10^\circ$ degrees away from the $\mathbf c^*$ axis and a broad minimum of less than 10\,T occurring near the $\mathbf{ab}$ plane. This result is in semiquantitative agreement with the theoretical expectation for the HK$\Gamma$ model with $\Gamma/K \simeq -0.5$~\cite{janssen2017}.

For in-plane fields, \rucl\ displays a combined two-fold and six-fold susceptibility anisotropy as a function of the field angle, as shown in Fig.~\ref{fig:rucl3-chi-phi}. The six-fold piece is consistent with the honeycomb structure and in agreement with the theoretical result for the HK$\Gamma$ model shown in Fig.~\ref{fig:hkgamma-chi}, while the two-fold piece arises from trigonal distortions \cite{kelley2018a} and, for low fields, from unequal zigzag domain population \cite{kelley2018b}.

\subsubsection[\nio]{\nio. }
In \nio, the susceptibility anisotropy is opposite to the one in \rucl, see Fig.~\ref{fig:rucl3-na2iro3-chi-T}(b): The in-plane susceptibility $\chi_{ab}$ is significantly smaller than the out-of-plane susceptibility $\chi_{c^*}$, with a ratio $\chi_{ab}/\chi_{c^*}$ ranging from about $0.6$ at room temperature to a minimum near the N\'eel temperature of about $0.4$ \cite{singh2010}. The largest contribution to this anisotropy most likely originates from an anisotropy in the $g$ tensor, which is expected to arise from a trigonal crystal field. This in turn is again consistent with the expectation that the intrinsic anisotropy arising from the off-diagonal Gamma interaction is significantly smaller in \nio\ than in \rucl.
In a finite magnetic field, the magnetization is largely featureless and increases linearly with field up to 60\,T; magnetic torque measurements show anomalies only in fields above 30\,T \cite{das_sebastian2019}. The latter have been interpreted as evidence for a dominant ferromagnetic Kitaev exchange interaction, in agreement with the estimates from ab-initio calculations~\cite{katukuri2014, yamaji2014, sizyuk2014, winter2016}.


\subsection{Field-induced phases and phase transitions in \rucl}
\label{subsec:rucl3-field}

For \rucl, it has been reported early on that the zigzag magnetic order can be suppressed by applying a moderate in-plane magnetic field, i.e., the N\'eel temperature drops to zero with increasing field \cite{johnson2015,sears2017,wolter2017}, see phase diagram in Fig.~\ref{fig:pd_rucl}. The corresponding critical field $\Hc$ is around 7--8\,T, its precise value depending on the in-plane field direction. In contrast, fields up to 30--40\,T applied perpendicular to the plane appear to leave the zigzag order intact.

For in-plane fields above $\Hc$ no thermal transition is observed, indicating a paramagnetic phase. Numerous observations such as NMR, thermodynamic measurements, and heat transport point to a sizable (spin) gap which closes at $\Hc$ and continuously reopens for $H>\Hc$, reaching 40--50\,K at 15\,T \cite{sears2017, baek2017}. We note that in specific-heat measurements, temperatures below 1\,K are required to see clear signatures of the gap for fields smaller than 12\,T, while the specific heat above 2\,K is consistent with approximate power-law behavior \cite{wolter2017, zheng2017}.
The transition at $\Hc$ has been analyzed in detail in Ref.~\cite{wolter2017}: Measurements down to 0.4\,K show signatures of an exponentially suppressed specific heat at low $T$, indicating an excitation gap both for $H<\Hc$ and $H>\Hc$, with the gap being minimal (or zero) at $\Hc \approx 7$\,T, see Fig.~\ref{fig:pd_rucl}. The specific-heat data display approximate scaling behavior signifying a quantum critical point at $\Hc$, with exponents consistent with the Ising universality class in $2+1$ dimensions.

\begin{figure*}[t]
\center{
\includegraphics[width=0.95\linewidth]{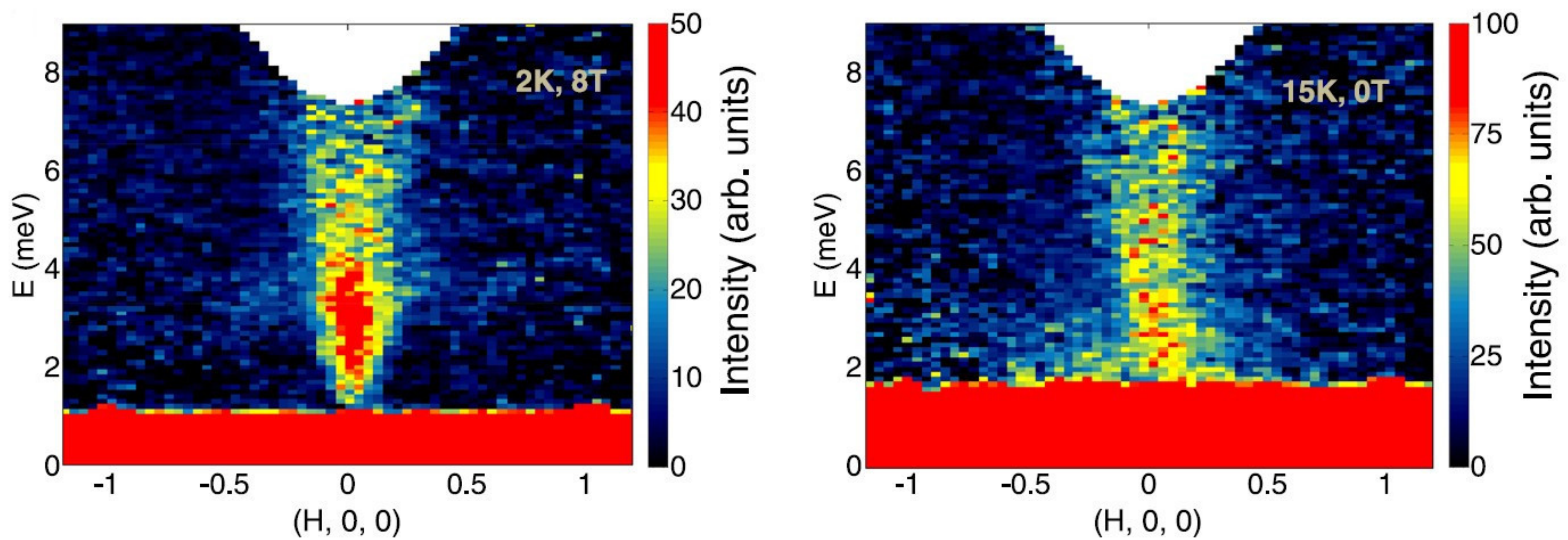}
}
\caption{
Inelastic neutron scattering intensity of \rucl\ in the paramagnetic phase as function of energy and wavevector along $(H,0,0)$ in the trigonal notation, corresponding to the $\mathbf b$ axis in Fig.~\ref{fig:rucl3}.
Left: $T=2$\,K and $\mu_0 H=8$\,T with $\mathbf H \parallel \mathbf a$, i.e., near the QPT at $\Hc$.
Right: $T=15$\,K and $\mu_0 H=0$\,T.
Reprinted from \cite{banerjee2018}.
}
\label{fig:banerjee}
\end{figure*}

The destruction of zigzag order at in-plane fields of about 7--8\,T is consistent with theoretical results for realistic HK$\Gamma$ models, treated either in the semiclassical limit \cite{janssen2017} or for $S=1/2$ \cite{winter2018} (with the precise value of $\Hc$ depending on model parameters including the $g$ tensor). The field-induced phase above $\Hc$ has been interpreted as a (topological) quantum spin liquid \cite{baek2017}. This would require the existence of another quantum phase transition at higher fields where the spin liquid would give way to the trivial high-field phase; for in-plane fields such a transition has not been clearly detected to date and also appears absent from numerical results.

A detailed analysis of AC susceptibility measurements at low $T$ did find \emph{two} transitions as function of applied in-plane field \cite{kelley2018b}. Here, the {\em upper} transition corresponds to the destruction of antiferromagnetic order at $\Hc\approx 7$\,T mentioned above, whereas the lower transition occurs at $\Hc'\approx 6$\,T within the ordered phase. Both $\Hc$ and $\Hc'$ show a characteristic dependence on the in-plane field direction, with sixfold periodicity. Hence, \rucl\ displays an intermediate ordered phase, and Ref.~\cite{kelley2018b} suggested that a canted antiferromagnetic state with larger intralayer unit cell may be realized for $\Hc'<H<\Hc$, the existence of which is triggered by the proximity to the Kitaev-Gamma model.

\subsubsection[Magnetic excitation spectrum]{Magnetic excitation spectrum.}

Inelastic neutron scattering has been employed to map out the magnetic excitations of \rucl\ as function of energy and momentum, both at zero field \cite{banerjee2016a, banerjee2016b, do2017} and up to the critical field \cite{banerjee2018}.
The results document highly unconventional behavior: The zero-field zigzag-ordered state displays the expected spin-wave modes near the $\mathbf M$ points at low energy, however, in addition a large scattering intensity is seen near the $\boldsymbol{\Gamma}$ point over a large energy range up to 7\,meV. Heating above the N\'eel temperature eliminates the spin-wave modes, while the $\boldsymbol\Gamma$ feature acquires more intensity. Given that this feature is broad in momentum space, it has been interpreted in terms of proximate spin-liquid behavior, and comparison with the spectrum of the Kitaev model \cite{knolle2014} has been made \cite{banerjee2016a,banerjee2016b}.

Remarkably, applying an in-plane magnetic field has an effect onto the excitation spectrum similar to raising the temperature: The low-temperature excitation spectrum at 8\,T, i.e., close to $\Hc$, also displays a strong signal centered at $\boldsymbol \Gamma$, but essentially no intensity near the $\mathbf M$ points, see Fig.~\ref{fig:banerjee}.
This is noteworthy in two aspects: First, little intensity near the $\mathbf M$ points implies that any critical spin dynamics near $\mathbf M$ -- as a precursor of the low-field ordered state -- must be restricted to energies below 1\,meV. Second, the spectrum indeed indicates close proximity of the system near $\Hc$ to a spin liquid.

The excitation spectrum at zero wavevector has also been probed using techniques of electron spin resonance, microwave, and THz absorption \cite{little2017, wang2017a, ponomaryov2017, kataev2018}. These results collectively show low-energy magnetic excitation modes both above and below $\Hc$. These modes soften and broaden upon approaching $\Hc$, indicating an excitation continuum. In addition, a background continuum reaching down to 0.4\,meV has been reported\cite{kataev2018} to exist over a broad field range below $\Hc$.

\begin{figure*}[t]
\includegraphics[width=\linewidth]{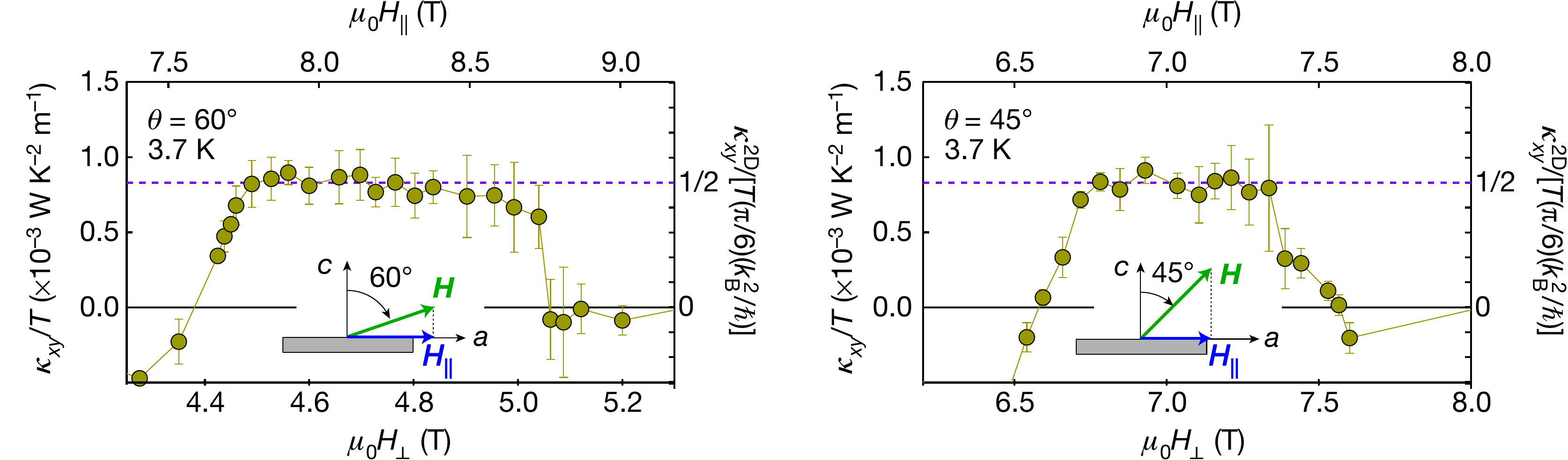}
\caption{
Transverse thermal conductivity of \rucl, plotted as $\kappa_{xy}/T$, as function of magnetic field which is titled by an angle $\theta$ away from the $\mathbf c^\ast$ axis towards the $\mathbf a$ axis. Over a range of fields $\kappa_{xy}/T$ is approximately quantized, corresponding to $\nth=1/2$, i.e., a single chiral Majorana edge mode, see text.
Left: $\theta=60^{\circ}$,
Right: $\theta=45^{\circ}$.
Reprinted from \cite{kasahara2018b}.
}
\label{fig:kasahara}
\end{figure*}

\subsubsection[Thermal transport and thermal Hall effect]{Thermal transport and thermal Hall effect.}

The longitudinal thermal conductivity, $\kappa_{xx}$ of \rucl\, as function of field has been measured in Refs.~\cite{leahy2017,yu2018,hentrich2018}. At low temperatures, the in-plane thermal conductivity shows a minimum around $\Hc$ and strongly increases for larger fields. A plausible explanation is that the thermal transport is phonon-dominated, with phonon scattering by magnetic excitations being strongest near $\Hc$ where the spin gap is small. Clear-cut signatures of magnetic heat transport have not been detected in $\kappa_{xx}$ to our knowledge.

Remarkably, clear indications of magnetic heat transport have been found, however, in the \emph{transverse} thermal conductivity $\kappa_{xy}$, i.e., the thermal Hall effect. While $\kappa_{xy}$ is finite in bulk metals and quantized in conventional quantum Hall states, $\kappa_{xy}/T = \nth [(\pi k_\mathrm{B}^2)/(6\hbar)]$, with $\nth$ being the number of heat-carrying chiral edge modes, $\kappa_{xy}$ in magnetic insulators is often essentially zero due to the absence of a Lorentz force. Exceptions to the latter are non-trivial chiral magnetic states. Therefore it came as a surprise when a significant $\kappa_{xy}$ was measured in \rucl\ for out-of-plane fields in the paramagnetic phase, i.e., above the N\'eel temperature \cite{kasahara2018a, hentrich2019}.

The perhaps most exciting observation concerns the finding of a half-quantized thermal Hall effect in Ref.~\cite{kasahara2018b}: At low temperatures around 3.5--5\,K and in a particular window of magnetic field a thermal Hall signal corresponding to $\nth=1/2$ was measured in \rucl, see Fig.~\ref{fig:kasahara}. This observation appears tied to a field titled 45$^\circ$ or 60$^\circ$ away from the $\mathbf c^\ast$ axis and emerges at field strengths of 9--10\,T. A thermal Hall effect with $\nth=1/2$ is expected from a chiral edge mode of Majorana fermions in the presence of a bulk gap, and this is indeed realized in the $B$ phase of the Kitaev model in an applied field, as discussed in Sec.~\ref{sec:th_kitaev}. Consequently, the experiment of Ref.~\cite{kasahara2018b} has been interpreted as direct evidence for a field-induced topological spin-liquid phase with a Majorana edge mode in \rucl. Subsequent theory work \cite{vinkler2018,ye2018} has argued that the expected quantization of $\kappa_{xy}/T$ \cite{kitaev2006} can survive even in the presence of strong phonon heat conductivity (i.e., small Hall angles). Further experiments are needed to determine the evolution of $\kappa_{xy}$ at temperature below 3.5\,K and to search for clear signatures of the quantum phase transition(s) bounding the topological phase.

\section{Outlook}   \label{sec:outlook}

Spin-orbit-coupled magnets in general and Kitaev materials in particular constitute a highly active field of condensed-matter research. In this review article, we have summarized the current understanding of the behavior of Kitaev magnets in external magnetic fields, covering both theoretical and experimental results. We have highlighted the strongly anisotropic magnetic responses, the occurrence of novel field-induced phases, and the possibility for topological magnon excitations.

Progress in the field can be expected along different avenues:
First, the synthesis and investigation of novel candidate materials will broaden the materials base. One interesting case in point is TbInO$_3$, realizing a spin-orbit-coupled honeycomb magnet with no detectable magnetic order at low temperatures~\cite{gaulin2019}. 
Recent suggestions for Kitaev materials also include YbCl$_3$, which exhibits an interesting field response~\cite{xing2019, luo2019}.
Second, careful studies of low-temperature thermodynamic and transport properties as function of continuous field strength \textit{and angle} are required to uncover the rich phenomenology expected on theoretical grounds.
Third, the progress of numerical methods for two-dimensional spin systems, most notably variants of density matrix renormalization group approaches, will yield a more comprehensive picture of quantum phase diagrams in applied fields. Fourth, conceptual and field-theoretic ideas will help to understand the quantum phase transitions observed both in experiment and numerics. Together, this will pave the way to applications involving, e.g., emergent Majorana-fermion modes.

\ack

The authors thank E.\ C.\ Andrade, C.\ Balz, W.\ Brenig, B.\ B\"uchner, L.\ Fritz, P.\ Gegenwart, D.\ G.\ Joshi, V.\ Kataev, W.\ G.\ F.\ Kr\"{u}ger, P.\ Lampen-Kelley, D.\ G.\ Mandrus, T.\ Meng, R.\ Moessner, S.\ E.\ Nagler, S.\ Rachel, A.\ Rosch, K.\ P.\ Schmidt, U.\ F.\ P.\ Seifert, R.\ Valent\'i, J.\ van den Brink, S.\ M.\ Winter, A.\ U.\ B.\ Wolter, and F.\ Zschocke for instructive discussions and for collaborations on related work. We are grateful to D.\ G.\ Joshi for useful comments on the manuscript.
This research was supported by the DFG through SFB 1143, GRK 1621, and the W\"urzburg-Dresden Cluster of Excellence on Complexity and Topology in Quantum Matter -- \textit{ct.qmat} (EXC 2147, project id 39085490).

\appendix
\section[]{Spin-wave theory in the high-field phase}   \label{app:spin-wave}

\paragraph{High-field limit.}
Spin-wave theory relies on a semiclassical expansion with the inverse spin magnitude $1/S$ serving as a formal control parameter, where $S \to \infty$ corresponds to the classical limit. For $S=1/2$, this is in general not a reliable approach if strong frustration is present. However, for field strengths $h$ that are much larger than the relevant exchange interactions, quantum fluctuations are suppressed for any fixed $S$, suggesting $1/h$ as an alternative control parameter of the semiclassical expansion.
Let us consider the HK$\Gamma$ model, with an additional third-neighbor Heisenberg interaction $J_3$, in an external field~$\mathbf h$,
\begin{eqnarray}
\mathcal H_\mathrm{HK\Gamma}^\mathrm{Z} & = & \sum_{\langle i j \rangle_\gamma} \left[
	J \mathbf S_i \cdot \mathbf S_j + K S_i^\gamma S_j^\gamma
	+ \Gamma (S_i^\alpha S_j^\beta + S_i^\beta S_j^\alpha)
	\right]
	\nonumber \\* &&
	+ \sum_{\lllangle ij \rrrangle} J_3 \mathbf S_i \cdot \mathbf S_j
	- \mathbf h \cdot \sum_i \mathbf S_i,
\end{eqnarray}
where $(\alpha, \beta, \gamma) = (x,y,z)$ on a $z$ bond and cyclically permuted on $x$ and $y$ bonds. 
Magnons are the collective excitations that describe the low-energy quantum fluctuations about the classical ground state. 
We consider a general field direction
\begin{equation}
	\mathbf h/h = \hat \mathbf a \sin \vartheta_h \cos \varphi_h + \hat \mathbf b \sin\vartheta_h \sin\varphi_h + \hat\mathbf c^* \cos\vartheta_h,
\end{equation}
where $\hat \mathbf a$ and $\hat \mathbf b$ are normalized crystal in-plane lattice vectors in the monoclinic basis and $\hat \mathbf c^*$ is the unit vector in the out-of-plane direction; cf.\ Fig.~\ref{fig:rucl3}. $\vartheta_h \in [0,\pi]$ and $\varphi_h \in [0,2\pi)$ denote the field's out-of-plane and in-plane angles, respectively.
In the high-field phase, the classical ground state is $\langle \mathbf S/S \rangle_{S \to \infty} = \mathbf n$ with
\begin{equation}
	\mathbf n = \hat \mathbf a \sin \vartheta_S \cos \varphi_S + \hat \mathbf b \sin\vartheta_S \sin\varphi_S + \hat\mathbf c^* \cos\vartheta_S.
\end{equation}
As discussed in Sec.~\ref{subsec:kitaev-gamma}, the spins' out-of-plane angle $\vartheta_S$ is in general \emph{not} equal to the field angle $\vartheta_h$ in the presence of a finite off-diagonal Gamma interaction. In the classical limit, the relation between $\vartheta_S$ and $\vartheta_h$ is given by the implicit equation \cite{janssen2017}
\begin{equation} \label{eq:theta-S}
	\frac{h}{S} \sin(\vartheta_h - \vartheta_S) + \frac{3\Gamma}{2} \sin 2\vartheta_S  = 0.
\end{equation}
Note that $\vartheta_S = 0$ ($\vartheta_S = \pi/2$) for $\vartheta_h = 0$ ($\vartheta_h = \pi/2$), but in general $\vartheta_S$ differs from $\vartheta_h$.
On the other hand, the in-plane angles are always equal, $\varphi_S=\varphi_h$, such that the projection of the spin onto the $ab$ plane is always parallel to the field's in-plane component.

\paragraph{Holstein-Primakoff bosons.}

In the spin-wave theory, each spin operator $\mathbf S_i$ is replaced by bosonic operators $a_i$ and $a_i^\dagger$, which satisfy the canonical commutation relations $[a_i,a_j^\dagger] = \delta_{ij}$ and $[a_i,a_j] = [a_i^\dagger,a_j^\dagger] = 0$. We employ a Holstein-Primakoff decomposition
\begin{eqnarray}
	S_i^+ & = & \sqrt{2S} \sqrt{1 - \frac{a_i^\dagger a_i}{2S}} a_i = \sqrt{2S} a_i + \mathcal O(1/\sqrt{S}), \label{eq:holstein-primakoff1}\\
	S_i^- & = & \sqrt{2S} a_i^\dagger \sqrt{1 - \frac{a_i^\dagger a_i}{2S}} = \sqrt{2S} a_i^\dagger + \mathcal O(1/\sqrt{S}), \label{eq:holstein-primakoff2}\\
	S_i^n & = & S - a_i^\dagger a_i. \label{eq:holstein-primakoff3}
\end{eqnarray}
Here, $S_i^n \equiv (\mathbf S \cdot \mathbf n)$ is the spin component along the classical spin direction $\mathbf n$ and $S_i^\pm \equiv (\mathbf S_i \cdot \mathbf e) \pm \iu [\mathbf S_i \cdot (\mathbf n \times \mathbf e)]$ are the ladder operators consisting of the orthogonal spin components, with $\mathbf e$ being an (arbitrary) unit vector perpendicular to $\mathbf n$.
In Eqs.~(\ref{eq:holstein-primakoff1}-\ref{eq:holstein-primakoff3}), the last expression respectively corresponds to the linear spin-wave theory, which leads to the noninteracting magnon picture.

A straightforward calculation leads to the quadratic Hamiltonian
\begin{eqnarray}
	\mathcal{H}_\mathrm{HK\Gamma}^\mathrm{Z} & = & S^2 \varepsilon_\mathrm{cl}
	\nonumber \\* && \hspace{-1em}
	+ \frac{S}{2} \sum_{\mathbf q}
	\left(\begin{array}{@{}c@{}}
	\vec \alpha_\mathbf{q} \\
	\vec \alpha^*_{-\mathbf{q}}
	\end{array}\right)^\dagger
	\left(\begin{array}{@{}cc@{}}
	K(\mathbf q) & \Delta^\dagger(\mathbf q) \\
	\Delta(\mathbf q) & K^\mathrm{T}(-\mathbf q)
	\end{array}\right)
	\left(\begin{array}{@{}c@{}}
	\vec \alpha_\mathbf{q} \\
	\vec \alpha^*_{-\mathbf{q}}
	\end{array}\right)
	\nonumber \\* && \hspace{-1em}
	+ \mathcal O(S^0),
\end{eqnarray}
where $S^2\varepsilon_\mathrm{cl}$ is the classical ground-state energy and
\begin{equation}
	\vec \alpha_{\mathbf{q}} =
	\left(\begin{array}{@{}c@{}}
	\sqrt{\frac{2}{N}}\sum_{i \in A} \mathrm{e}^{-\iu \mathbf q \cdot \mathbf R_i} a_i  \\
	\sqrt{\frac{2}{N}}\sum_{j \in B} \mathrm{e}^{-\iu \mathbf q \cdot \mathbf R_j} a_j
	\end{array}\right)	
\end{equation}
and $\vec \alpha_{\mathbf q}^* \equiv \left(\vec \alpha_{\mathbf q}^{\mathrm{T}}\right)^\dagger$ are the vectors of magnon annihilation and creation operators on the two sublattices $A$ and $B$ of the honeycomb lattice, with $\mathbf R_i$ and $\mathbf R_j$ corresponding to the respective position vectors and $N$ the number of sites. In the high-field phase of the HK$\Gamma$ model, the $2\times 2$ matrices $K(\mathbf q)$ and $\Delta(\mathbf q)$ have the form
\begin{equation}
	K(\mathbf q) =
	\left(\begin{array}{@{}cc@{}}
	\varepsilon_0 & \lambda_0(\mathbf q) \\
	\lambda_0^*(\mathbf q) & \varepsilon_0
	\end{array}\right)
\end{equation}
and
\begin{equation}
	\Delta(\mathbf q) =
	\left(\begin{array}{@{}cc@{}}
	0 & \lambda_1(\mathbf q) \\
	\lambda_1(-\mathbf q) & 0
	\end{array}\right).
\end{equation}
In terms of the nearest-neighbor vectors $\boldsymbol\delta_x$, $\boldsymbol\delta_y$, and $\boldsymbol\delta_z$ on the $x$, $y$, and $z$ bonds of the honeycomb lattice and the cubic spin-space basis vectors $\mathbf e_x$, $\mathbf e_y$, $\mathbf e_z$ (cf.\ Fig.~\ref{fig:rucl3}), one finds for the off-diagonal entries
\begin{eqnarray}
	\lambda_0(\mathbf q) & = & \sum_\gamma \mathrm e^{\iu\mathbf q\cdot \boldsymbol\delta_\gamma}  \biggl\{ 
	J
	+ \frac{K}{2} \left[(\mathbf e \cdot \mathbf e_\gamma)^2 + ((\mathbf n \times \mathbf e) \cdot \mathbf e_\gamma)^2 \right]
	\nonumber \\* && \hspace{-2.5em}
	+ \Gamma \bigl[(\mathbf e \cdot \mathbf e_\alpha)(\mathbf e \cdot \mathbf e_\beta) + ((\mathbf n \times \mathbf e) \cdot \mathbf e_\alpha) ((\mathbf n \times \mathbf e) \cdot \mathbf e_\beta) \bigr]
	\biggr\}
	\nonumber \\* && \hspace{-2.5em}
	+ \sum_\gamma   \mathrm e^{-2\iu\mathbf q\cdot\boldsymbol\delta_\gamma} J_3
\end{eqnarray}
and
\begin{eqnarray}
	\lambda_1(\mathbf q) & = & \sum_\gamma  \mathrm e^{\iu\mathbf q\cdot\boldsymbol\delta_\gamma} \biggl\{
	\frac{K}{2} \left[(\mathbf e \cdot \mathbf e_\gamma) - \iu ((\mathbf n \times \mathbf e) \cdot \mathbf e_\gamma) \right]^2
	\nonumber \\* && \hspace{-2.5em}
	+ \Gamma \bigl[(\mathbf e \cdot \mathbf e_\alpha)(\mathbf e \cdot \mathbf e_\beta) - ((\mathbf n \times \mathbf e) \cdot \mathbf e_\alpha) ((\mathbf n \times \mathbf e) \cdot \mathbf e_\beta) 
	\nonumber \\* && \hspace{-2.5em}
		- \iu (\mathbf e \cdot \mathbf e_\alpha)((\mathbf n \times \mathbf e) \cdot \mathbf e_\beta) - \iu (\mathbf e \cdot \mathbf e_\beta)((\mathbf n \times \mathbf e) \cdot \mathbf e_\alpha) 
	\bigr]
	\biggr\}.
	\nonumber \\*[-0.25\baselineskip] &&
\end{eqnarray}
The diagonal entries in $K(\mathbf q)$ are $\mathbf q$ independent and read
\begin{eqnarray}
	\varepsilon_0 & = & - 3 J - 3 J_3 - K - 2 \Gamma \sum_\gamma (\mathbf n \cdot \mathbf e_\alpha) (\mathbf n \cdot \mathbf e_\beta)
	\nonumber \\* && 
	+ \frac{h}{S} \cos(\vartheta_h - \vartheta_S).
\end{eqnarray}
The diagonal entries in $\Delta(\mathbf q)$ become finite in the presence of a second-neighbor Kitaev interaction \cite{janssen2017}.
We note that $\mathcal O(\sqrt{S})$ contributions to $\mathcal H_\mathrm{HK\Gamma}^\mathrm{Z}$ cancel if and only if $\vartheta_S$ satisfies Eq.~(\ref{eq:theta-S}).

\paragraph{Bosonic Bogoliubov transformation.}

The quadratic Hamiltonian in the linear spin-wave theory can be diagonalized by means of a bosonic Bogoliubov transformation~\cite{mucciuolo2004, wessel2005},
\begin{eqnarray}
	\left(\begin{array}{@{}cc@{}}
	\Omega(\mathbf q) & 0 \\
	0 & \Omega(\mathbf q)
	\end{array}\right)
	& = &
	T^\dagger(\mathbf q)
	\left(\begin{array}{@{}cc@{}}
	K(\mathbf q) & \Delta^\dagger(\mathbf q) \\
	\Delta(\mathbf q) & K^\mathrm{T}(-\mathbf q)
	\end{array}\right)
	T(\mathbf q)
	\nonumber \\*[-0.25\baselineskip] &&
\end{eqnarray}
where $\Omega(\mathbf q) = \diag{\omega^{(1)}_{\mathbf q}, \omega^{(2)}_{\mathbf q}}$ and the transformation matrix can be written as
\begin{equation}
	T(\mathbf q) = 
	\left(\begin{array}{@{}cc@{}}
	U(\mathbf q) & V(\mathbf q) \\
	V^*(-\mathbf q) & U^*(-\mathbf q)
	\end{array}\right).
\end{equation}
The latter satisfies the orthogonality relations $T \Sigma T^\dagger = T^\dagger \Sigma T = \Sigma$, where $\Sigma = \diag{\mathds 1, -\mathds 1}$.
The spectrum of the Hamiltonian is given by the eigenvalue equation
\begin{eqnarray}
	\left(\begin{array}{@{}cc@{}}
	K(\mathbf q) & \Delta^\dagger(\mathbf q) \\
	-\Delta(\mathbf q) & -K^\mathrm{T}(-\mathbf q)
	\end{array}\right)
	\left(\begin{array}{@{}c@{}}
	\vec u^{(n)}_\mathbf{q} \\
	\vec v^{*(n)}_{-\mathbf q}
	\end{array}\right)	
	& = & 
	\omega^{(n)}_\mathbf{q}
	\left(\begin{array}{@{}c@{}}
	\vec u^{(n)}_\mathbf{q} \\
	\vec v^{*(n)}_{-\mathbf q}
	\end{array}\right).
	\nonumber \\*[-0.25\baselineskip] && 
\label{eq:eigenvectors-bogoliubov}
\end{eqnarray}
If the eigenvector $|n(\mathbf q) \rangle \equiv \left(\vec u^{(n)}_\mathbf{q}, \vec v^{*(n)}_{-\mathbf q}\right)^\mathrm{T}$ is normalized with respect to the inner product involving the matrix~$\Sigma$, i.e., $\langle n(\mathbf q) | \Sigma | n(\mathbf q)\rangle = 1$ with $\langle n(\mathbf q)| \equiv | n(\mathbf q)\rangle^\dagger$, then the columns of the matrix $T(\mathbf q)$ are given by the two vectors $|n(\mathbf q) \rangle$.
Note that the $4\times 4$ matrix in the above equation is non-Hermitian, but all four eigenvalues are real. They come in pairs of opposite signs and the physical spectrum consists of those two that are positive and for which the corresponding transformation matrix $T$ satisfies the orthogonality relations \cite{mucciuolo2004}.

\paragraph{Magnetization correction.}

In the cases when $\langle \mathbf S_i \rangle \parallel \mathbf h$, the $1/S$ correction to the classical magnetization is simply given by the magnon density $\langle a_i^\dagger a_i\rangle$,%
\footnote{In the cases when $\langle \mathbf S_i \rangle \nparallel \mathbf h$ in the classical limit, i.e., $\vartheta_S \neq \vartheta_h$, one also needs to take the quantum correction to $\vartheta_S - \vartheta_h$ into account.}
\begin{eqnarray}
	\frac{m_\parallel}{S} 
	& = &
	\frac{1}{N} \sum_i \left(1 - \frac{1}{S} \langle a_i^\dagger a_i \rangle \right) + \mathcal O(1/S^2) \\
	& = & 
	1 - \frac{1}{2S} \sum_{n=1,2} \int \frac{d^2\mathbf q}{(2\pi)^2} \left|\vec v_{-\mathbf q}^{*(n)}\right|^2 + \mathcal O(1/S^2),
	\nonumber \\*[-0.5\baselineskip] &&
\end{eqnarray}
where $\vec v_{-\mathbf q}^{*(n)}$ denotes the lower half of the normalized $n$-th eigenvector occurring in Eq.~(\ref{eq:eigenvectors-bogoliubov}), with positive energy $\omega_\mathbf{q}^{(n)}$. The momentum integral is over all wavevectors $\mathbf q = (q_x, q_y)$ in the Brillouin zone.

\paragraph{Dynamic spin structure factor.}

The dynamic spin structure factor $\mathcal S(\mathbf q, \omega)$ reflects the time-dependent spin correlations and is in principle experimentally accessible by inelastic neutron or X-ray scattering. In linear spin-wave theory, it can be computed for finite frequency $\omega > 0$ in terms of the two normalized eigenvectors $\left(\vec u^{(n)}_\mathbf{q}, \vec v^{*(n)}_{-\mathbf q}\right)^\mathrm{T}$ as
\begin{eqnarray}
	\mathcal S(\mathbf q, \omega) & = & \frac{1}{N} \sum_{i,j = 1}^{N} \mathrm{e}^{\mathrm{i} \mathbf{q} \cdot (\mathbf R_i - \mathbf R_j)}  
	\int_{-\infty}^\infty dt \mathrm{e}^{\mathrm{i}\omega t} \langle \mathbf{S}_i(t) \cdot \mathbf{S}_j(0) \rangle	
%
%
	\nonumber \\ & = &
	\frac{S}{2} \sum_{n=1}^2 \sum_{m,m'=1}^{2} 2\pi \delta(\omega-\omega_{\mathbf q}^{(n)})
	\nonumber \\ && \times
	 \left[ u_{\mathbf{q}m}^{(n)} u_{\mathbf{q} m'}^{*(n)}  + v_{-\mathbf{q}m}^{*(n)} v_{-\mathbf{q} m'}^{(n)} \right] + \mathcal O(\delta(\omega), S^0),
	\nonumber \\*[-0.5\baselineskip] &&
\end{eqnarray}
where $u_{\mathbf{q}m}^{(n)}$ and $v_{-\mathbf{q}m}^{*(n)}$ are the $m$-th components of $\vec u^{(n)}_\mathbf{q}$ and $\vec v^{*(n)}_{-\mathbf q}$, respectively.

\paragraph{Chern number.}

The topological character of the $n$-th magnon band can be computed in terms of the Chern number \cite{thouless1982, berry1984}
\begin{equation}
	C^{(n)} = \frac{1}{2\pi\iu} \int d^2\mathbf q F^{(n)}_{xy} (\mathbf q),
\end{equation}
where the Berry curvature is $F^{(n)}_{xy}(\mathbf q) = \frac{\partial  A_y(\mathbf q)}{\partial q_x} - \frac{\partial A_x(\mathbf q)}{\partial q_y}$ and the Berry connection is
\begin{equation}
	(A_x^{(n)}(\mathbf q), A_y^{(n)}(\mathbf q)) = \langle n(\mathbf q)| \Sigma \nabla_{\mathbf q} | n(\mathbf q) \rangle,
\end{equation}
with the eigenvector $|n (\mathbf q) \rangle$ again normalized with respect to the matrix $\Sigma$. A discrete version of the above definition is given in Ref.~\cite{joshi2018}.

\paragraph{Thermodynamic observables.}

Thermodynamic and transport quantities in the high-field phase are given in terms of the spectrum $\omega_\mathbf{q}^{(n)}$ and the Berry curvature $F^{(n)}_{xy}(\mathbf q)$.
The specific heat, for instance, can be written as \cite{wolter2017}
\begin{equation}
	C_\mathrm{mag}(T) = \sum_{n=1,2} \int \frac{d^2\mathbf{q}}{(2\pi)^2} \frac{\partial}{\partial T} \frac{\omega^{(n)}_\mathbf{q}}{\mathrm{e}^{\omega^{(n)}_\mathbf{q} / (k_\mathrm B T)} - 1}.
\end{equation}
The thermal Hall conductivity is given by \cite{matsumoto2011, mcclarty2018}
\begin{eqnarray}
	\frac{\kappa_{xy}}{T} & = & \frac{k_\mathrm{B}^2}{\iu\hbar} \sum_{n=1,2} \int d^2\mathbf{q} 
	c_2([\mathrm{e}^{\omega_\mathbf{q}^{(n)}/(k_\mathrm B T)} - 1]^{-1}) F_{xy}^{(n)}(\mathbf q) ,
	\nonumber \\*[-0.5\baselineskip] &&
\end{eqnarray}
where $c_2(\rho) = \int_0^{\rho} dt \ln^2(1 + t^{-1})$.

\section*{References}

\end{document}